\documentclass[superscriptaddress,aps,preprint,amsmath,amssymb,floatfix]{revtex4}

\usepackage{graphicx,morefloats}
\usepackage{subfigure} 
\usepackage{color,soul}
\usepackage{bm}
\usepackage{amsmath}
\usepackage{amssymb}
\usepackage{times}
\usepackage{hyperref}
\usepackage{bbm}

\newcommand{\SBP}[1]{\color{blue}{#1} \color{black}}

\begin{document}


\hyphenation{va-ni-sh-ing}

\begin{center}

\thispagestyle{empty}

{\large\bf Pristine quantum criticality in a Kondo semimetal}
\\[0.3cm]

W.\ T.\ Fuhrman$^{1,\oplus,\dagger}$, A. Sidorenko$^{2,\oplus}$,  J.\
H\"anel$^2$, H.\ Winkler$^2$, A.\ Prokofiev$^2$, J.\ A.\
Rodriguez-Rivera$^{3,4}$, Y. Qiu$^4$, P.\ Blaha$^5$, Q.\ Si$^6$, C. L.
Broholm$^{1,4,7}$, and S.~Paschen$^{2,6,\ast}$\\[0.3cm]

$^1$Institute for Quantum Matter and Department of Physics and Astronomy, The Johns Hopkins University, Baltimore, Maryland 21218 USA\\[-0.0cm]

$^2$Institute of Solid State Physics, Vienna University of Technology, Wiedner Hauptstr. 8-10, 1040
Vienna, Austria\\[-0.0cm]

\centerline{$^3$Department of Materials Sciences, University of Maryland, College Park, Maryland 20742, USA}

$^4$NIST Center for Neutron Research, National Institute of Standards and Technology, Gaithersburg, MD 20899, USA\\[-0.cm]

$^5$Institute of Materials Chemistry, Vienna University of Technology, 1040 Vienna, Austria\\[-0.cm]

$^6$Department of Physics and Astronomy, Rice Center for Quantum Materials, Rice University, Houston, Texas 77005, USA\\[-0.cm]

$^7$Department of Materials Science and Engineering, The Johns Hopkins University, Baltimore, Maryland 21218 USA

\end{center}

\vspace{-0.15cm}

{\bf The observation of quantum criticality in diverse classes of strongly
correlated electron systems has been instrumental in establishing ordering
principles, discovering new phases, and identifying the relevant degrees of
freedom and interactions. At focus so far have been insulators and metals.
Semimetals, which are of great current interest as candidate phases with
nontrivial topology, are much less explored in experiments. Here we study the
Kondo semimetal CeRu$_4$Sn$_6$ by magnetic susceptibility, specific heat, and
inelastic neutron scattering experiments. The power-law divergence of the
magnetic Gr\"unesien ratio reveals that, surprisingly, this compound is quantum
critical without tuning. The dynamical energy over temperature scaling in the
neutron response, seen throughout the Brillouin zone, as well as the temperature
dependence of the static uniform susceptibility indicate that temperature is the
only energy scale in the criticality. Such behavior, which has been associated
with Kondo destruction quantum criticality in metallic systems, may well be
generic in the semimetal setting.}
\vspace{0.15cm}

\noindent E-mail: $^{\ast}$paschen@ifp.tuwien.ac.at
\vspace{-0.1cm}

\noindent $^{\oplus}$W.T.F and A.S. contributed equally to this work.
\vspace{-0.1cm}

\noindent $^{\dagger}$Present addresses: The Johns Hopkins Applied Physics Laboratory

\newpage

Quantum criticality is observed in many strongly correlated materials classes,
with quantum spin systems \cite{Col10.1}, high-$T_{\rm{c}}$ cuprate
\cite{Sac10.1} and iron pnictide \cite{Dai09.1} superconductors, and heavy
fermion metals \cite{Loe07.1,Kir20.1} being prominent examples. Among them, the
quantum critical insulators are the best understood. For instance, in the
insulating quantum magnet LiHoF$_4$ (Ref.\,\citenum{Bit96.1}), the
experimentally detected quantum criticality is well described in terms of the
Landau framework, i.e., by the critical fluctuations of the magnetic order
parameter \cite{Her76.1,Mil93.1}. In quantum critical metals, by contrast, the
underlying physics is much richer. In some systems the Landau description works
well \cite{Kue03.1,Jar10.1}, but in others it appears to fail
\cite{Sch00.1,Kir20.1}. Theoretically, the charge carriers introduce new
couplings to the order parameter or its underlying building blocks, which raises
new possibilities for quantum criticality \cite{Si01.1,Lee09.2,Mro10.1}.

\begin{figure*}
\vspace{-1.2cm}

\hspace{-15.3cm}{\bf\textsf{a}}
\vspace{-0.2cm}

\includegraphics[width=0.9\textwidth]{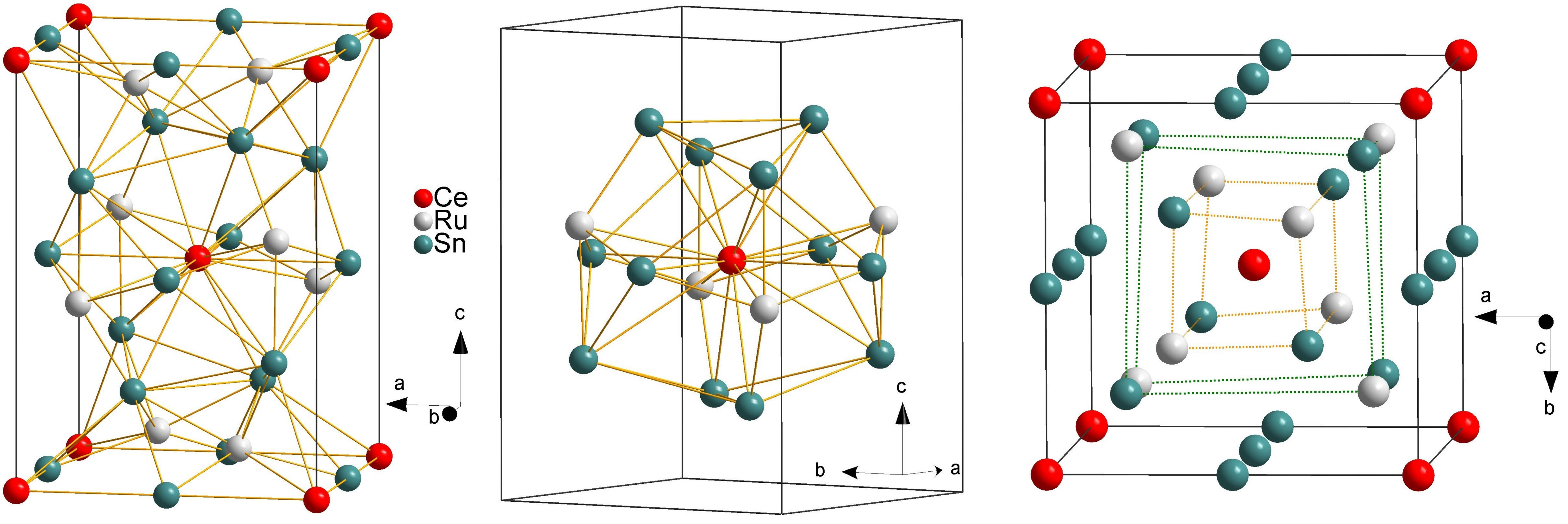}
\vspace{0.5cm}

\includegraphics[height=0.6\textwidth]{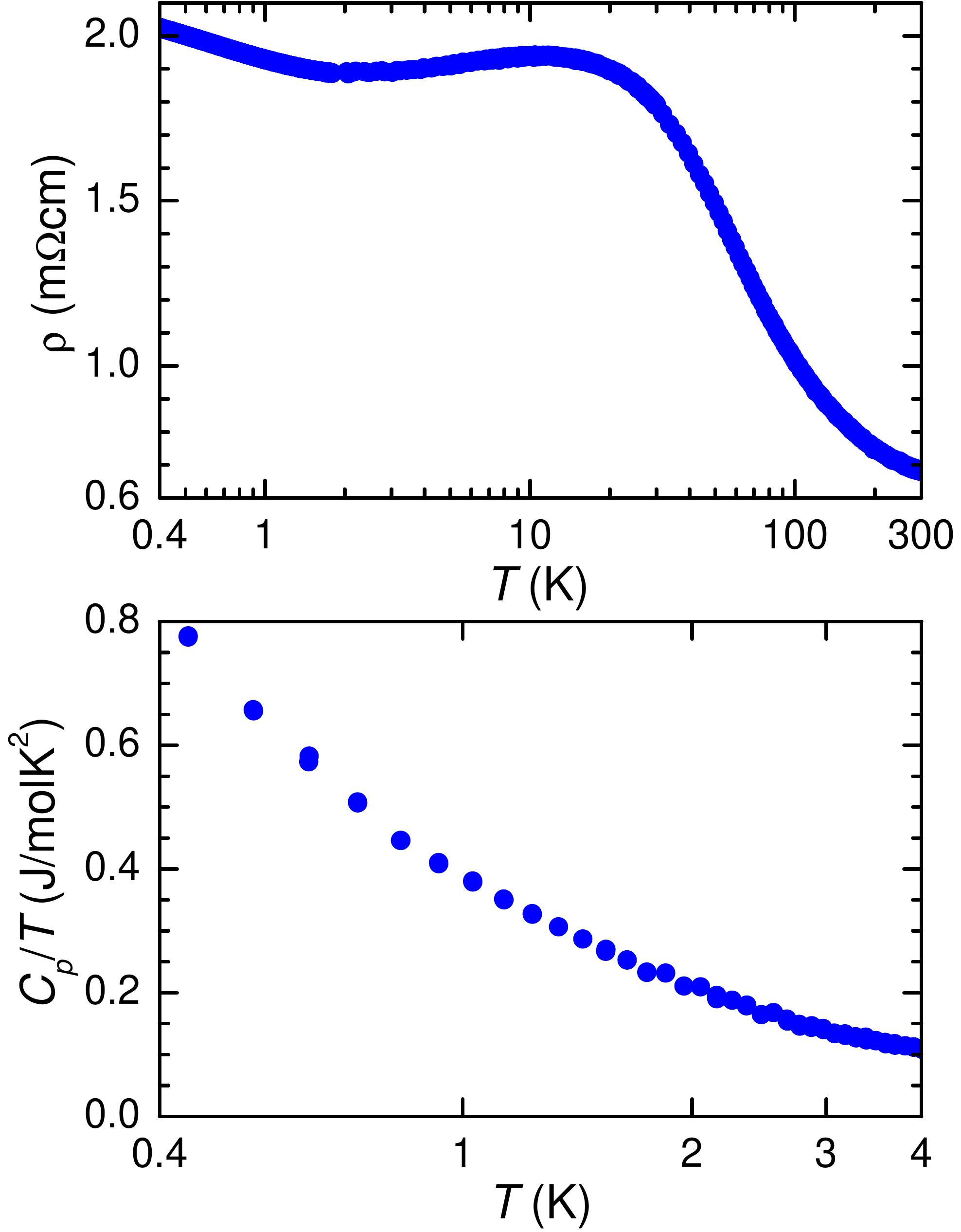}\hspace{0.2cm}\includegraphics[height=0.6\textwidth]{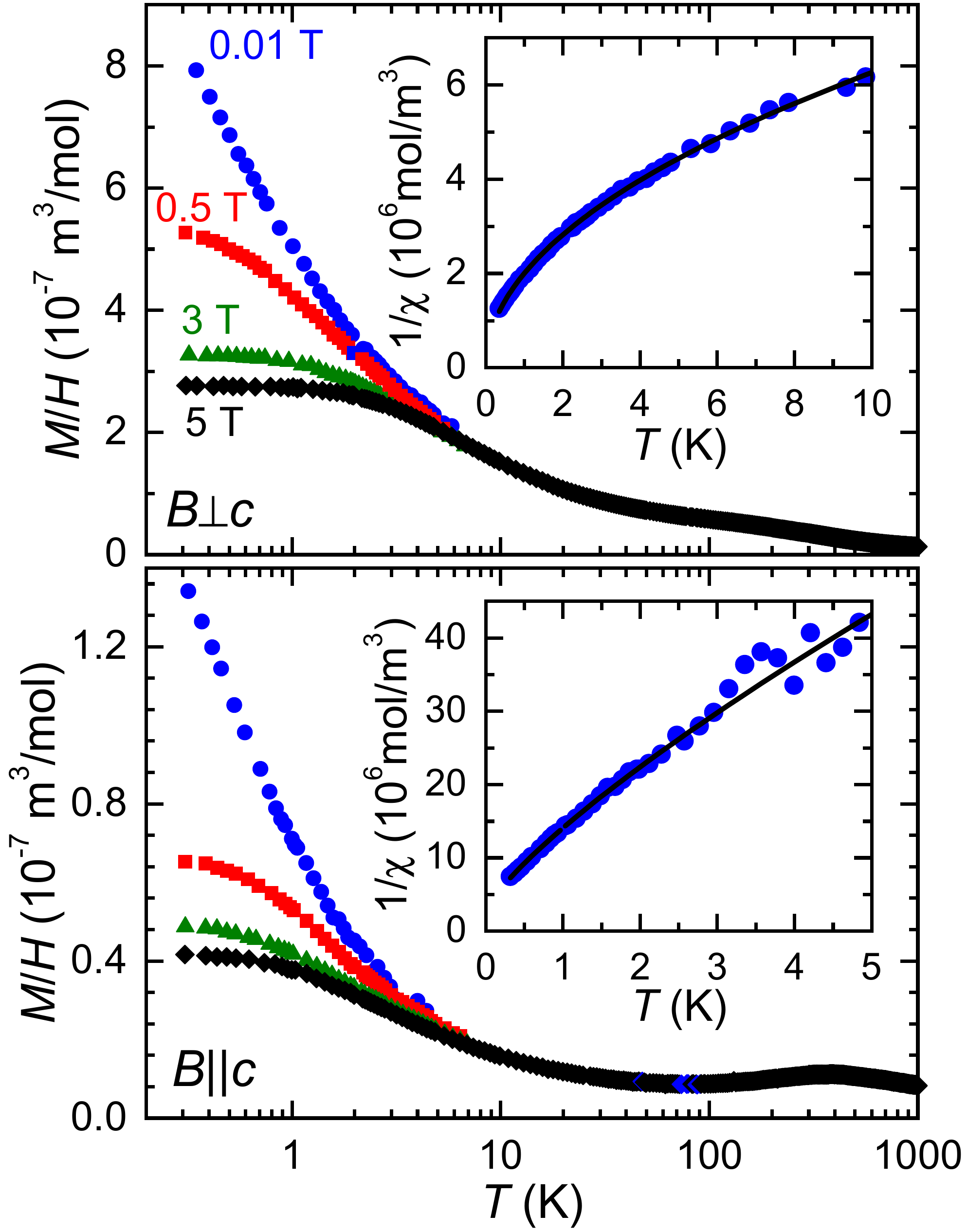}
\vspace{-10.6cm}

\hspace{-7.5cm}{\bf\textsf{b}} \hspace{7.4cm} {\bf\textsf{d}}
\vspace{4cm}

\hspace{-7.5cm}{\bf\textsf{c}} \hspace{7.4cm} {\bf\textsf{e}}
\vspace{5cm}

\caption{\label{fig1} {\bf Characterization of CeRu$_4$Sn$_6$.} {\bf (a)}
Unit cell with view along tetragonal direction $a=b$ (left), polyhedron around central Ce atom (center), and unit cell with view along $c$ of the noncentrosymmetric tetragonal $I{\bar 4}2m$ crystal structure of CeRu$_4$Sn$_6$,
with $a = 6.8810$\,\AA\ and $c = 9.7520$\,\AA\ (Refs.\citenum{Ven90.1,Das92.1}).
{\bf (b)} Electrical resistivity vs temperature, with current within the
tetragonal plane, evidencing the semimetallic character of CeRu$_4$Sn$_6$. {\bf
(c)} Low-temperature specific heat over temperature ratio, plotted on
semi-logarithmic scales, evidencing a stronger than logarithmic divergence. {\bf
(d)} Magnetization over magnetic field, measured in different magnetic fields ($B=\mu_0 H$)
applied perpendicular the tetragonal $c$ axis, as function of temperature. The
inset shows the inverse susceptibility, defined as $1/\chi = H/M|_{\rm 10\,mT}$, with a least-square
fit with $1/\chi = 1/\chi_0 +c\cdot T^{\alpha}$ to the date below 10\,K, with $\alpha = 0.50 \pm 0.01$. {\bf (e)} Same as {\bf d} for magnetic fields
applied along $c$, with $\alpha = 0.78 \pm 0.03$. The fit was done below 3\,K, where the noise in the data is small (note that $c$ is the hard direction, along which $\chi$ is much smaller).}

\end{figure*}

Here we explore the case of the Kondo semimetal CeRu$_4$Sn$_6$. Its
noncentrosymmetric crystal structure (Fig.\,\ref{fig1}a), strong spin-orbit
coupling associated with the three large $Z$ elements, and the fact that the
anisotropy in its electronic dispersion \cite{Gur13.1} cannot be attributed to a
nodal ground state wave function of the Ce$^{3+} 4f^1$ electron \cite{Sun15.2}
have led to speculations \cite{Sun15.2} that the material may be topologically
nontrivial. Indeed, density functional theory (DFT) calculation within the local
density approximation (LDA) plus the Gutzwiller scheme, predict the material to
host Weyl nodes \cite{Xu17.1}. If the nodal excitations persist in a full
treatment of the Kondo effect, CeRu$_4$Sn$_6$ will be a Weyl-Kondo semimetal
\cite{Dzs17.1,Lai18.1,Dzs18.1x}. Our comprehensive investigation of the
magnetization, Gr\"uneisen parameter, and inelastic neutron scattering as
functions of temperature and magnetic field reveals that this material is
quantum critical without tuning, behavior that is only rarely observed
\cite{Mat11.1}. This raises the exciting question as to whether a Weyl-Kondo
semimetal phase may indeed be located nearby, nucleating out of the quantum
critical fluctuations.

The semimetallic character of CeRu$_4$Sn$_6$ is evidenced by the weak temperature
dependence of the electrical resistivity below about 30\,K (Fig.\,\ref{fig1}b). We have also measured the temperature
dependence of the specific heat $C_{p}$ (Fig.\,\ref{fig1}c) and the
magnetization $M$, the latter with different magnetic fields $H$ applied along
the two main crystallographic directions (perpendicular and parallel to $c$,
Fig.\,\ref{fig1}d and e, respectively). The high-temperature anisotropy was
recently shown to be due to single ion crystal field effects on the Ce$^{3+}
4f^1$ electrons \cite{Amo18.1}. Deviations from this behavior below room
temperature indicate a partial gapping of the electronic density of states
and/or the onset of Kondo screening \cite{Amo18.1}. Below 10\,K, a strong field
dependence is observed. The temperature dependence of the magnetic
susceptibility $\chi = M/H|_{\rm 10\,mT}$ is not of simple Curie-Weiss type
[$\chi = C/(T - \Theta)$] but, instead, is well described by
\begin{equation}
\frac{1}{\chi(T)} = \frac{1}{\chi(T=0)} + \frac{(aT)^\alpha}{c}\;.
\end{equation}
with $\alpha = 0.50 \pm 0.01$ and $\alpha = 0.78 \pm 0.03$ for magnetic fields
applied perpendicular and parallel to $c$, respectively (Fig.\,\ref{fig1}d,e
insets). This is inconsistent with quantum criticality involving a Lorentzian
fluctuation spectrum, where $\alpha =1$ (Curie-Weiss law) is expected. The
application of larger magnetic fields gradually restores Fermi liquid behavior,
i.e., a temperature-independent low-temperature magnetization
(Fig.\,\ref{fig1}d,e main panels).

\begin{figure*}
\begin{center}
\includegraphics[height=0.6\textwidth]{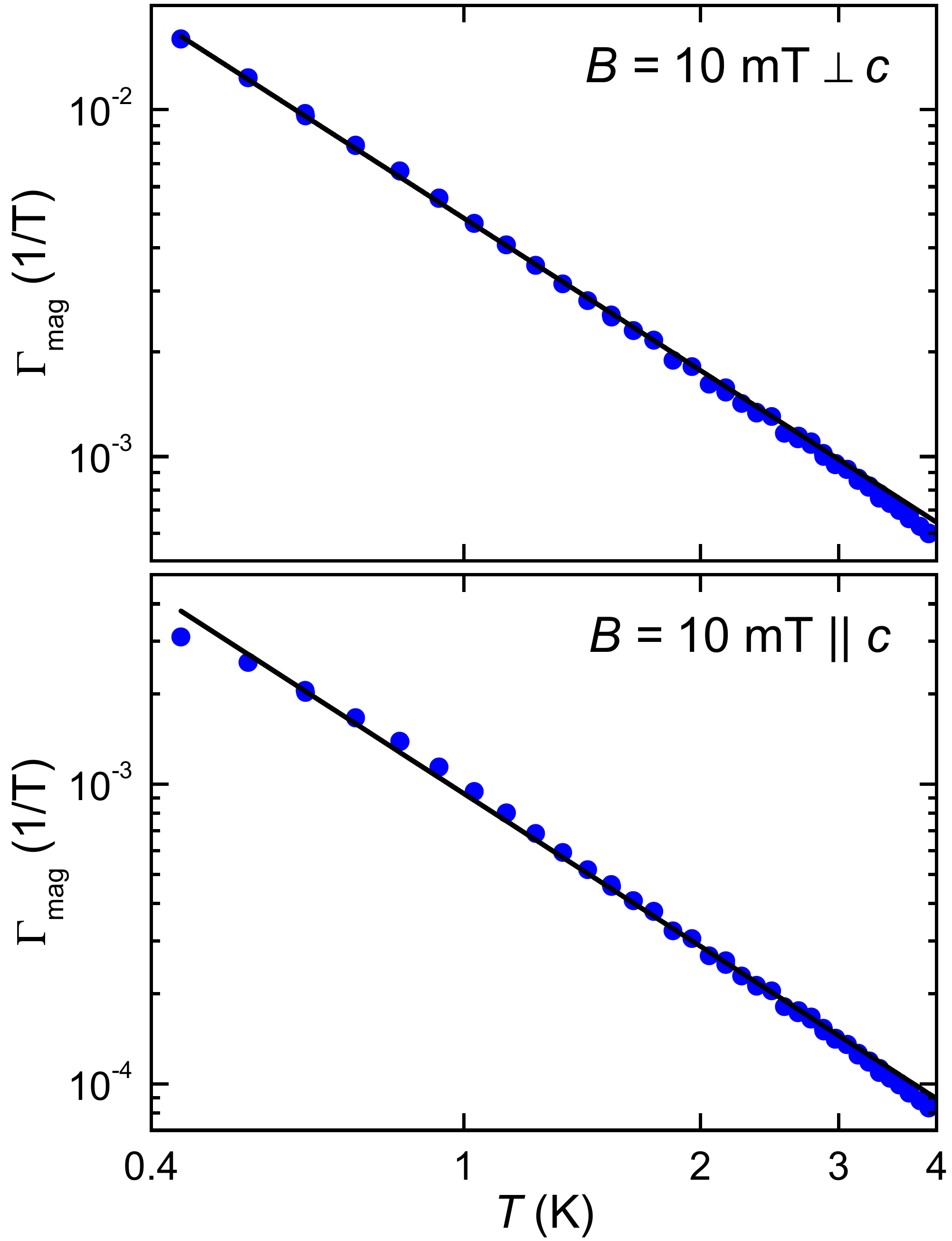}\hspace{0.6cm}\includegraphics[height=0.6\textwidth]{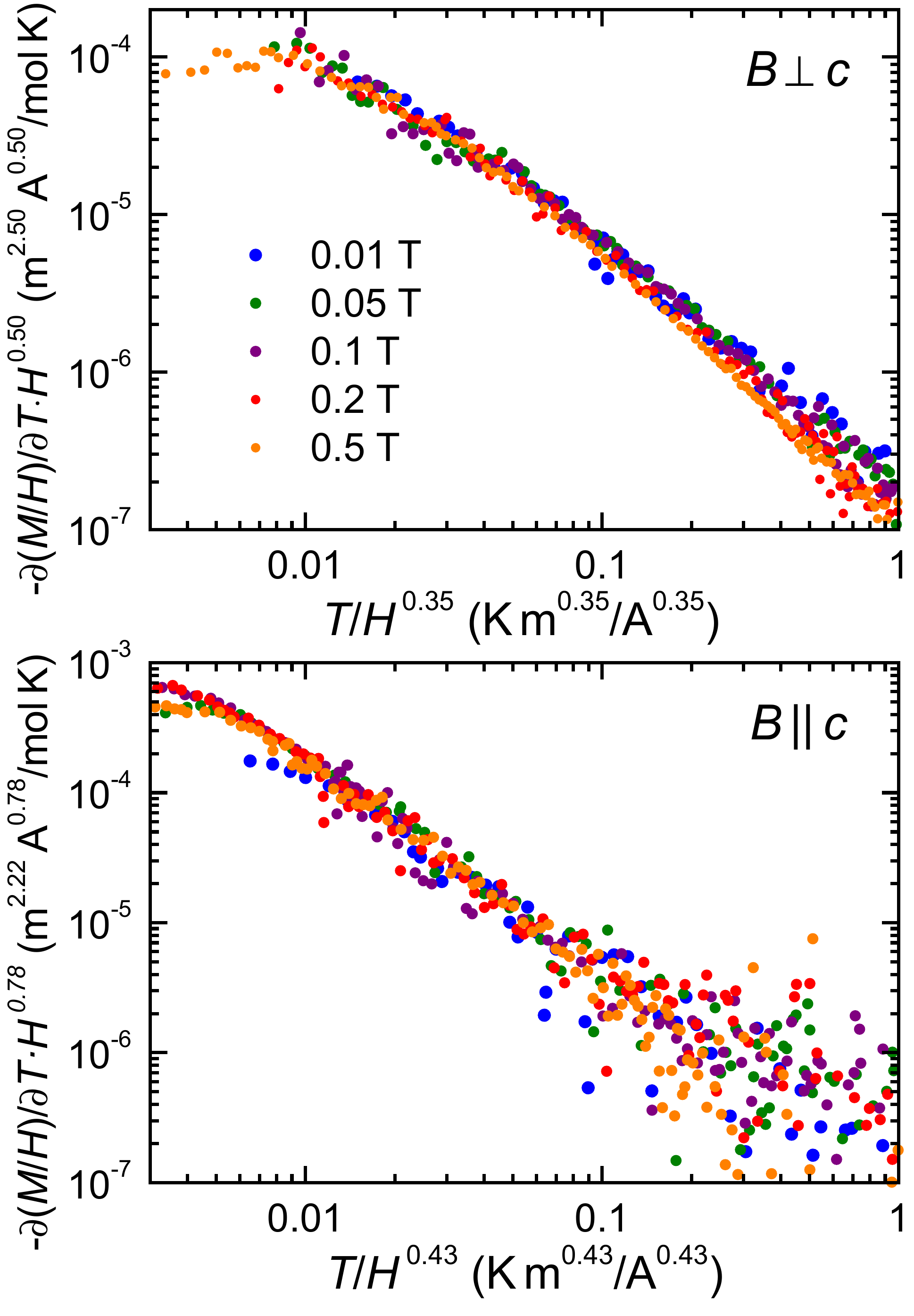}
\vspace{-10.6cm}

\hspace{-7cm}{\bf\textsf{a}} \hspace{7.4cm} {\bf\textsf{c}}
\vspace{3.8cm}

\hspace{-7cm}{\bf\textsf{b}} \hspace{7.4cm} {\bf\textsf{d}}
\vspace{6cm}

\caption{\label{fig2}
{\bf Signatures of quantum criticality in thermodynamic properties of
CeRu$_4$Sn$_6$.} {\bf (a)} Magnetic Gr\"uneisen ratio $\Gamma_{\rm{mag}}$ vs
temperature at $B=\mu_0 H =10$\,mT, applied perpendicular to the $c$ axis, with a power law
fit to the data (exponent $1.43 \pm 0.07$). {\bf (b)} Same as {\bf a} for
magnetic fields applied along $c$, with exponent $1.62 \pm 0.11$. {\bf (c)}
Scaled negative temperature derivative of the magnetization over field ratio, 
$-\partial(M/H)/\partial T\cdot H^{\beta}$ vs scaled temperature,
$T/H^{\gamma}$, for fields perpendicular to $c$, with exponents $\beta$ and $\gamma$ that collapse the data. {\bf (d)} Same as {\bf c} for
magnetic fields applied along $c$.}
\end{center}
\end{figure*}

The magnetic Gr\"uneisen ratio $\Gamma_{\rm{mag}} = -(\partial M/\partial
T)/C_{p}$ is expected to diverge at any quantum critical point \cite{Zhu03.1},
as observed in a number of quantum critical heavy fermion metals
\cite{Tok09.2,Geg16.1}. For CeRu$_4$Sn$_6$, from the low-field (10\,mT)
magnetization and specific heat data presented above, we find
$\Gamma_{\rm{mag}}\sim T^{-\epsilon}$ between about 0.4 and 4\,K, with $\epsilon
= 1.43 \pm 0.07$ and $1.62 \pm 0.11$ for fields perpendicular and parallel to
the $c$ axis, respectively (Fig.\,\ref{fig2}a,b), providing strong evidence for
quantum criticality without tuning in CeRu$_4$Sn$_6$.

The field and temperature scaling of the magnetization data underpins this
assignment. In Fig.\,\ref{fig2}c,d we plot  $-\partial(M/H)/\partial T\cdot
H^{\beta}$ vs $T/H^{\gamma}$ over more than two orders of magnitude in
$T/H^{\gamma}$, for fields between 10\,mT and 0.5\,T both perpendicular and
parallel to $c$. We find a good data collapse with the exponents $\beta$ being
equal to $\alpha$ extracted from Fig.\,\ref{fig1}d,e (insets) and $\gamma =
0.35 \pm 0.02$ and $0.43 \pm 0.02$ for fields perpendicular and parallel to $c$, respectively. The
fitted exponents are internally consistent, as they satisfy a scaling
relationship (see Supplementary Information). This kind of critical scaling,
with a fractional exponent $\alpha < 1$, indicates that the system is at, or
very close to, a beyond-Landau quantum critical point, and that magnetic field
acts as a tuning parameter, with criticality at $H=0$.

\begin{figure*}
\begin{center}
\hspace{-1.5cm}

\includegraphics[height=.28\textwidth]{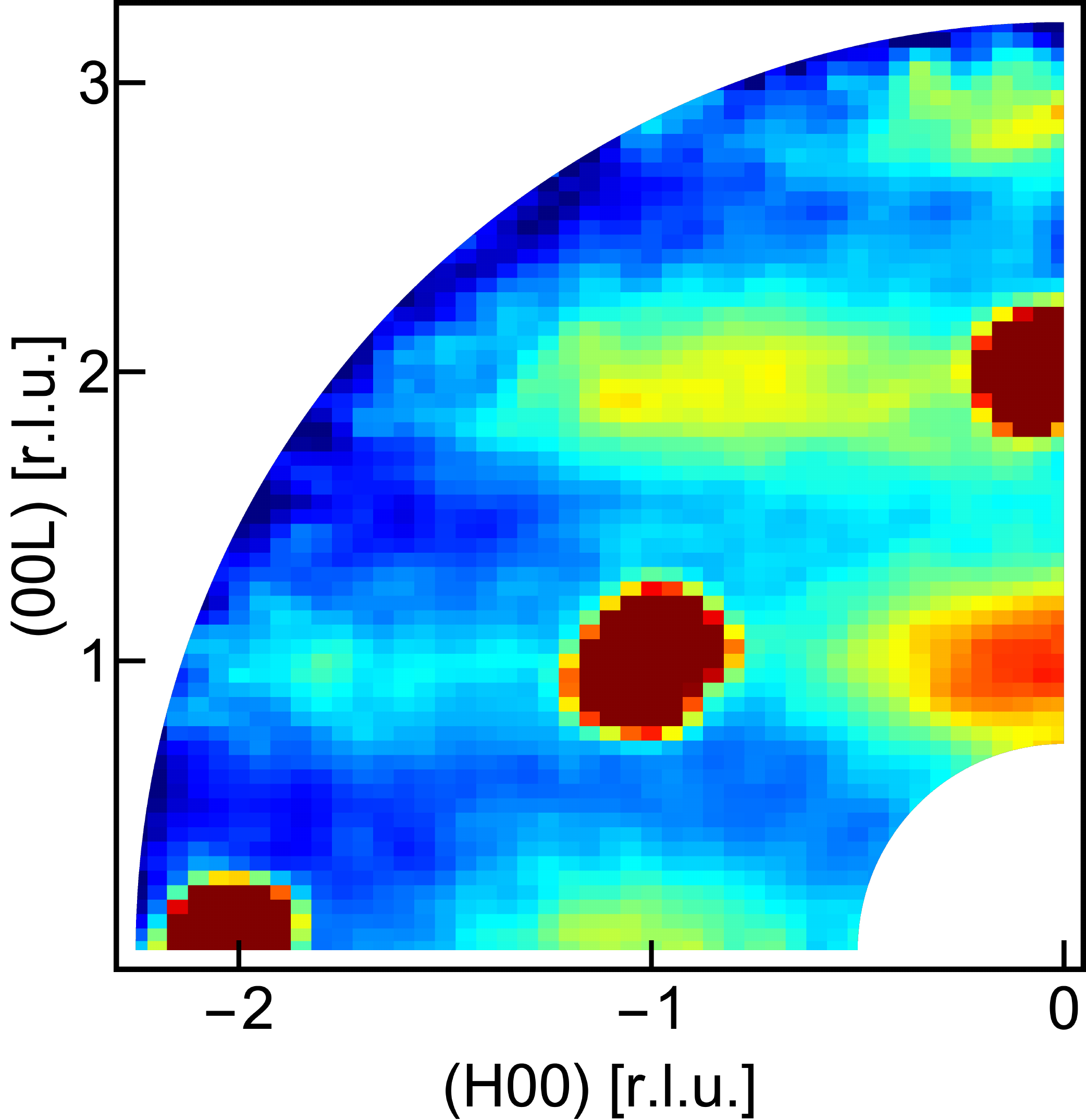}\hspace{0.3cm}\includegraphics[height=.28\textwidth]{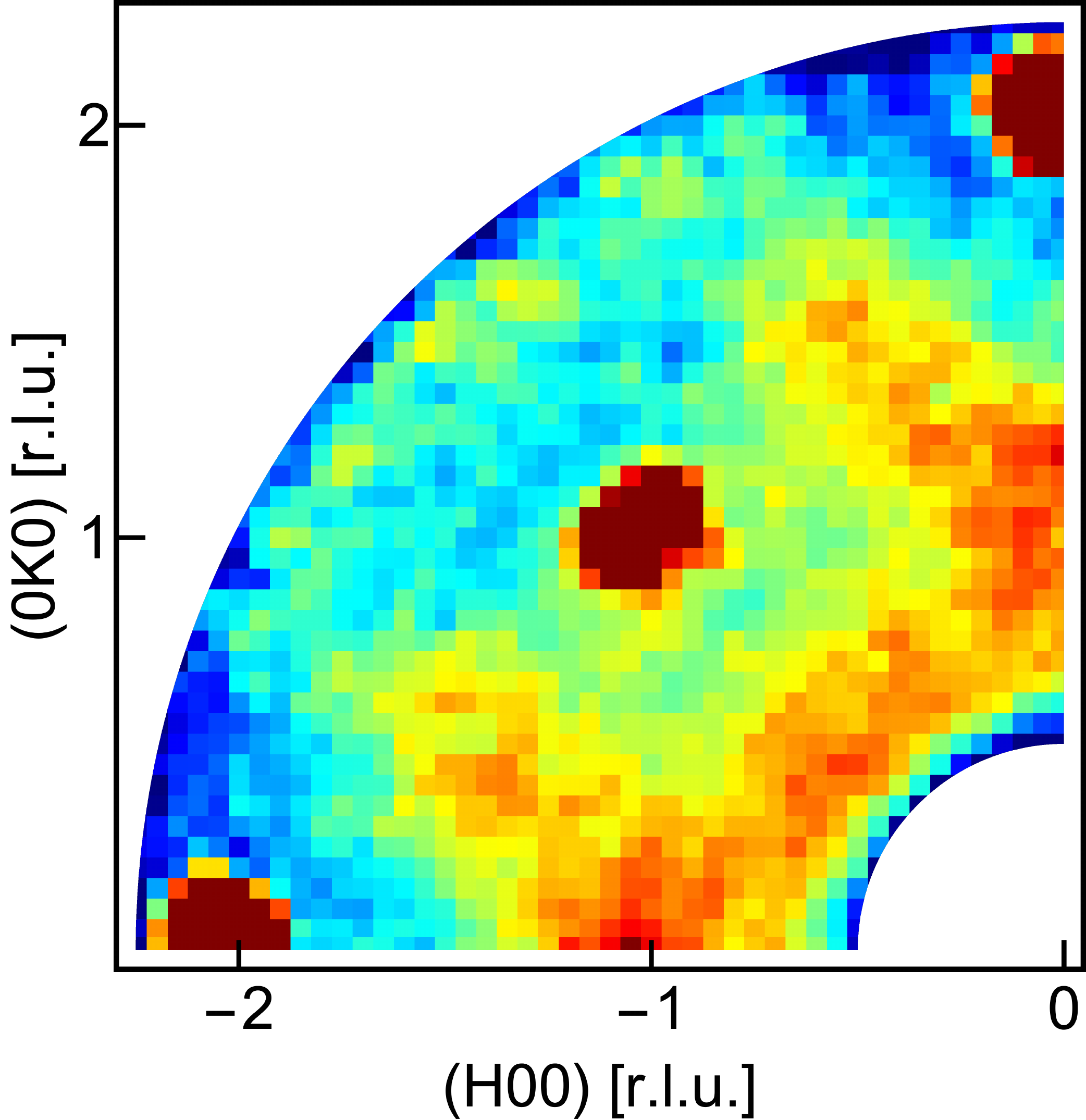}\hspace{0.3cm}\includegraphics[height=.28\textwidth]{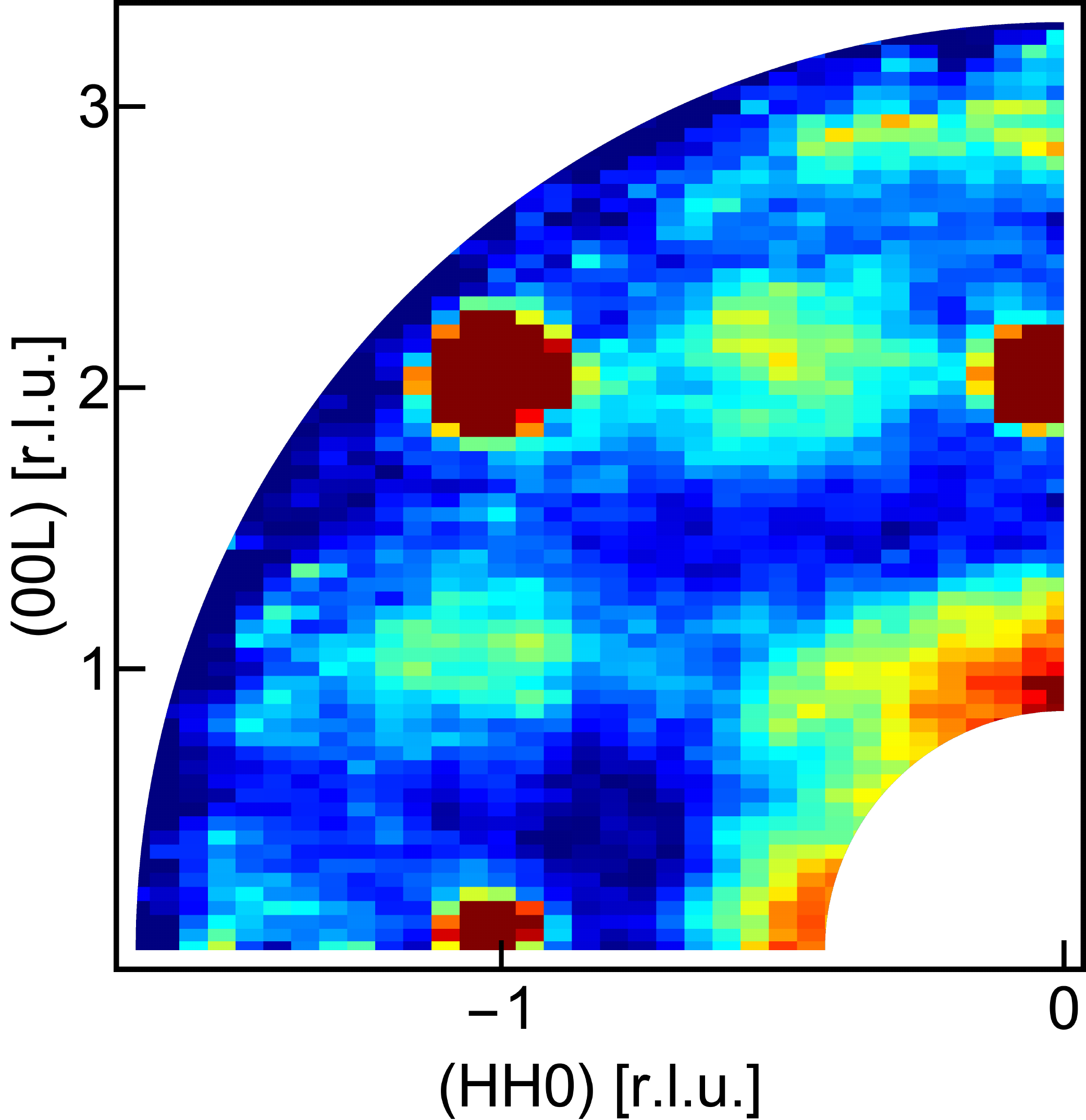}\hspace{0.3cm}\includegraphics[height=.3\textwidth]{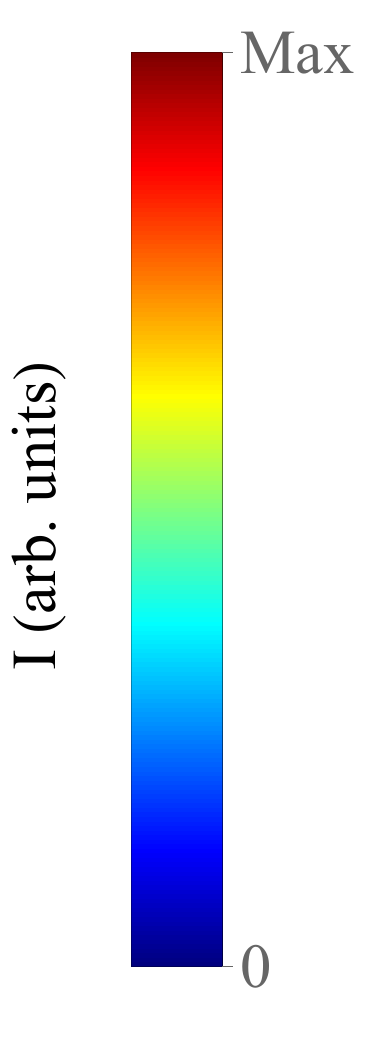}
\vspace{0.5cm}

\includegraphics[height=.26\textwidth]{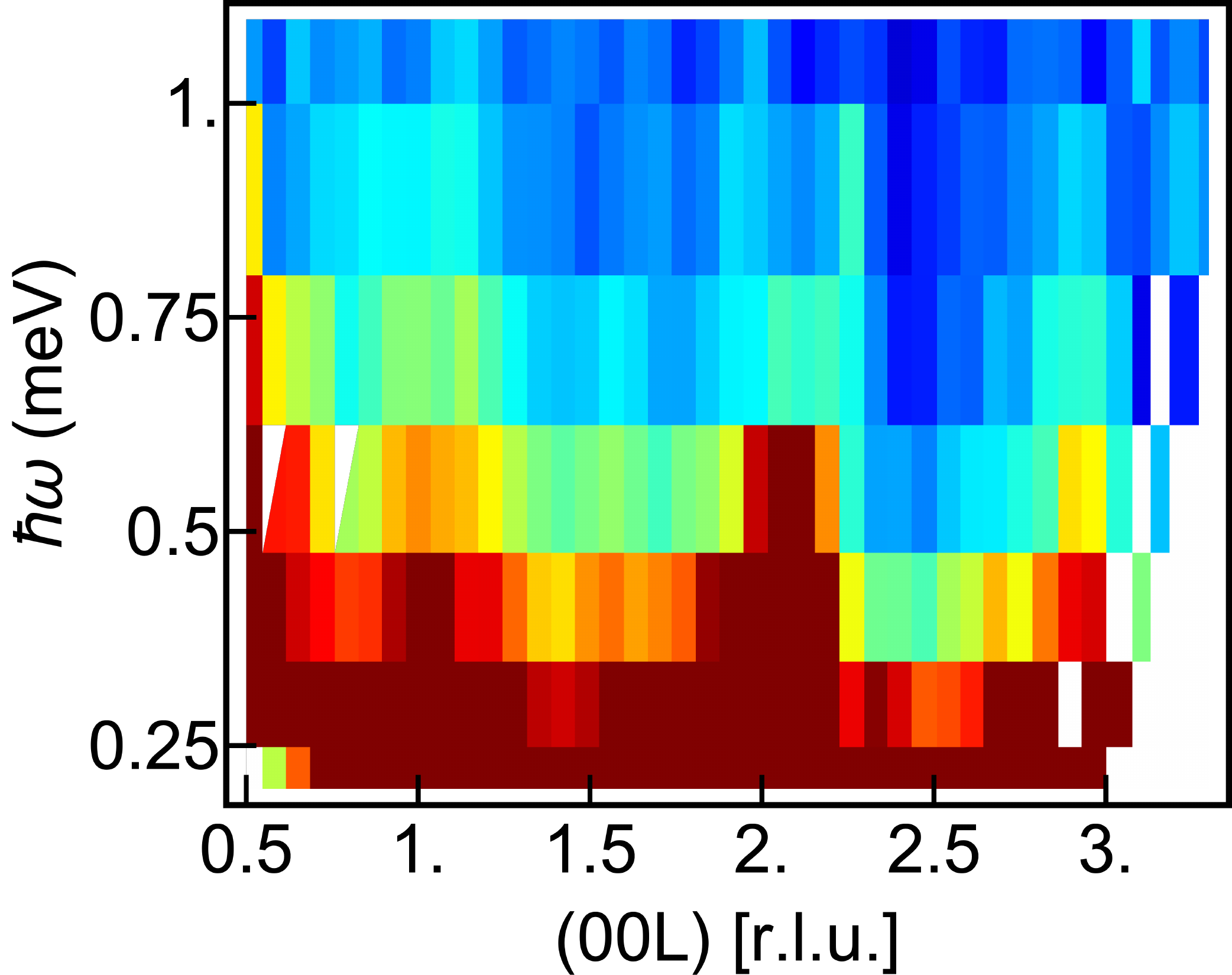}\hspace{0.3cm}\includegraphics[height=.26\textwidth]{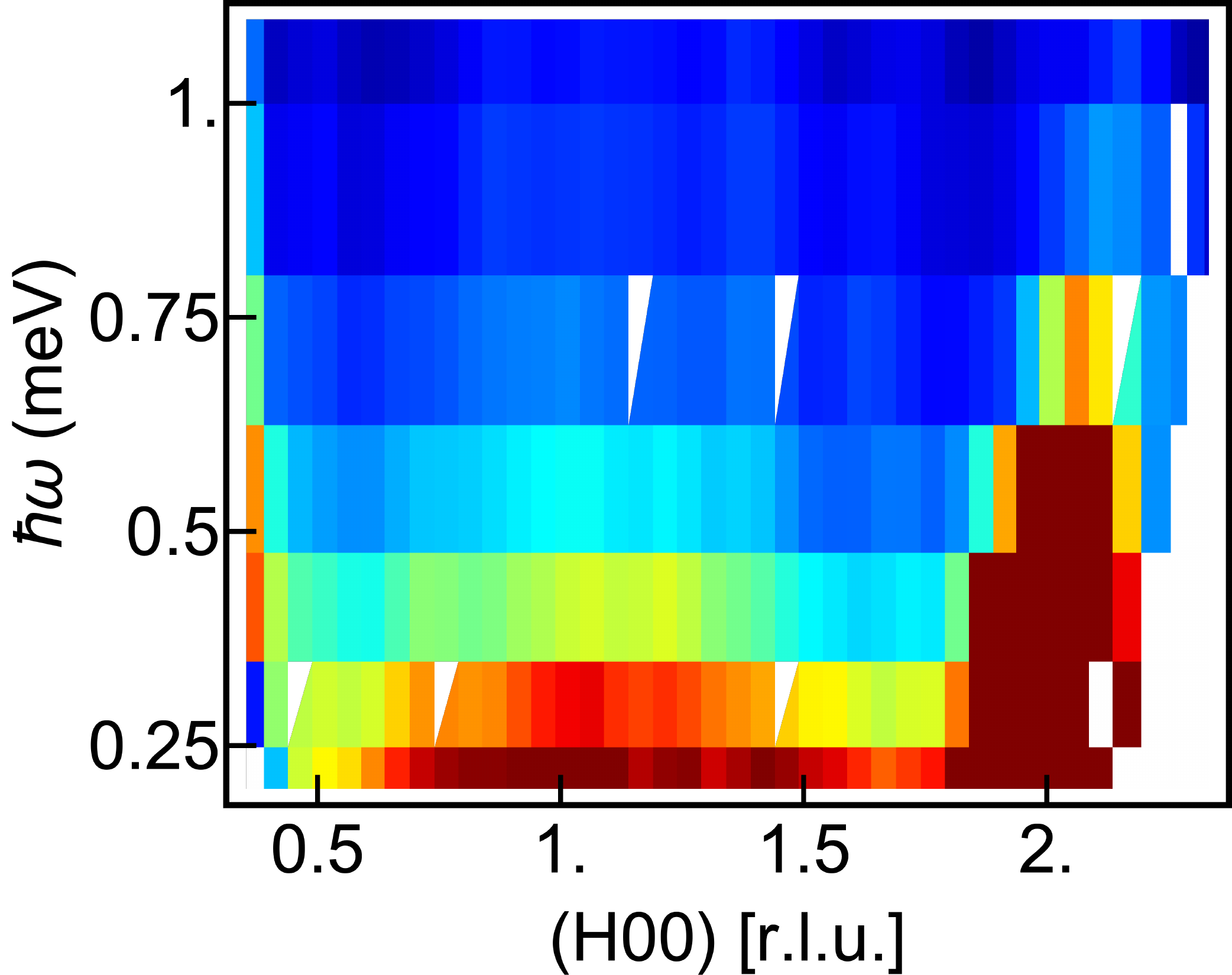}\hspace{0.2cm}\includegraphics[height=.26\textwidth]{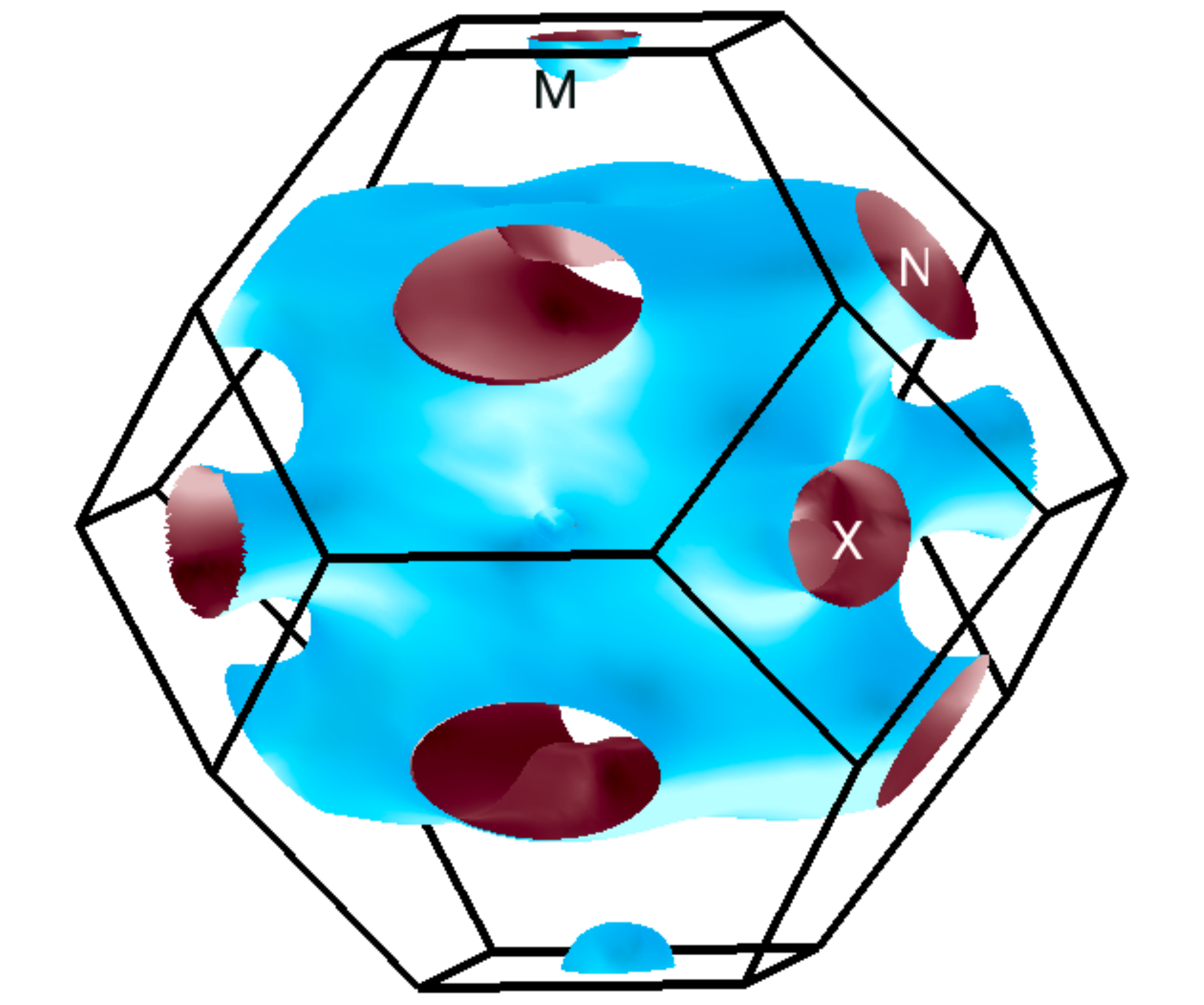}

\vspace{-10.2cm}

\hspace{-6.4cm}{\bf\textsf{a}}\hspace{4.55cm}{\bf\textsf{b}}\hspace{4.55cm}{\bf\textsf{c}}
\vspace{4.4cm}

\hspace{-4.6cm}{\bf\textsf{d}}\hspace{5.5cm}{\bf\textsf{e}}\hspace{5.35cm}{\bf\textsf{f}}
\vspace{5cm}

\caption{\label{fig3}
{\bf Quasielastic neutron scattering from CeRu$_4$Sn$_6$.} {\bf (a,b,c)} Energy
integrated neutron scattering intensity in the $({\rm H0L})$, $({\rm HK0})$, and
$({\rm HHL})$ planes, respectively. Saturated intensity at ${\rm
H\!\!+\!\!K\!\!+\!\!L\!=\!even}$ is residual Bragg intensity. Along
$(\frac{1}{2}\!\!+\!\!{\rm H}\, \frac{1}{2}\!\!-\!\!{\rm H}\, 0)$ and equivalent
directions in {\bf b} and, to a lesser extent, along  $({\rm H} 0 0)$ centered
at ${\rm H\!\!+\!\!0\!\!+\!\!L\!=\!odd}$ in {\bf a} the scattering intensity is
almost structureless. {\bf (d,e)} Energy vs momentum along $(0 0 {\rm L})$ and
$({\rm H} 0 0)$, respectively. The $Q$ and $\omega$ dependence of the scattering
cross section factorizes so that there is no dispersion and the signal continues
below the resolution limit of the instrument. {\bf (f)} ``$f$-core'' Fermi
surface derived from LDA-DFT calculations.}
\end{center}
\end{figure*}

This is corroborated by our neutron scattering investigation presented next. The
inelastic neutron scattering intensity is dominated by features broad in
momentum space, with no apparent energy scale, setting the lower bound for a gap
in the spin excitations to less than 0.1\,meV. In Fig.\,\ref{fig3}a-c we show
the intensity distribution at 1.5\,K, integrated from 0.2\,meV to 1.2\,meV, in the
$({\rm H}0{\rm L})$, $({\rm HK}0)$, and $({\rm HHL})$ planes, respectively. The energy
profile along two high symmetry directions, $(0 0 {\rm L})$ and  $({\rm H} 0
0)$, integrated over $\pm 0.2$ reciprocal lattice units (r.l.u.) perpendicular
to these directions,  is presented in Fig.\,\ref{fig3}d,e. The strong peaks at
${\rm H\!\!+\!\!K\!\!+\!\!L\!=\!even}$ are tails of nonmagnetic elastic Bragg scattering. There is
no apparent change in the momentum dependence with energy transfer, meaning that
the $Q$ and $\omega$ dependence of the scattering cross section factorizes and
there is no dispersion. This is clearly at odds with any description of the
criticality in terms of the fluctuations of an incipient symmetry-breaking order
parameter.  The broadness of the features is especially pronounced along
$(\frac{1}{2}\!\!+\!\!{\rm H}\, \frac{1}{2}\!\!-\!\!{\rm H}\, 0)$ (Fig.\,\ref{fig3}b) and, to a
lesser extent, along the $({\rm H} 0 0)$ direction for peak intensities at
${\rm H\!+\!\!0\!\!+\!\!L\!=\!odd}$ (Fig.\,\ref{fig3}a) and symmetry-related directions
where the scattering intensity simply follows the magnetic form factor of the
Ce$^{3+} 4f^1$ electrons. Scattering that is momentum-independent along lines in
the Brillouin zone is a sign of local (or Kondo destruction) quantum criticality
\cite{Si01.1} and has also been observed in quantum critical
CeCu$_{5.9}$Au$_{0.1}$ (Ref.\citenum{Sch00.1}). The wavevector dependence of the
scattering intensity away from these high-intensity lines is discussed below.

\begin{figure}

\begin{center}
\includegraphics[width=.72\textwidth]{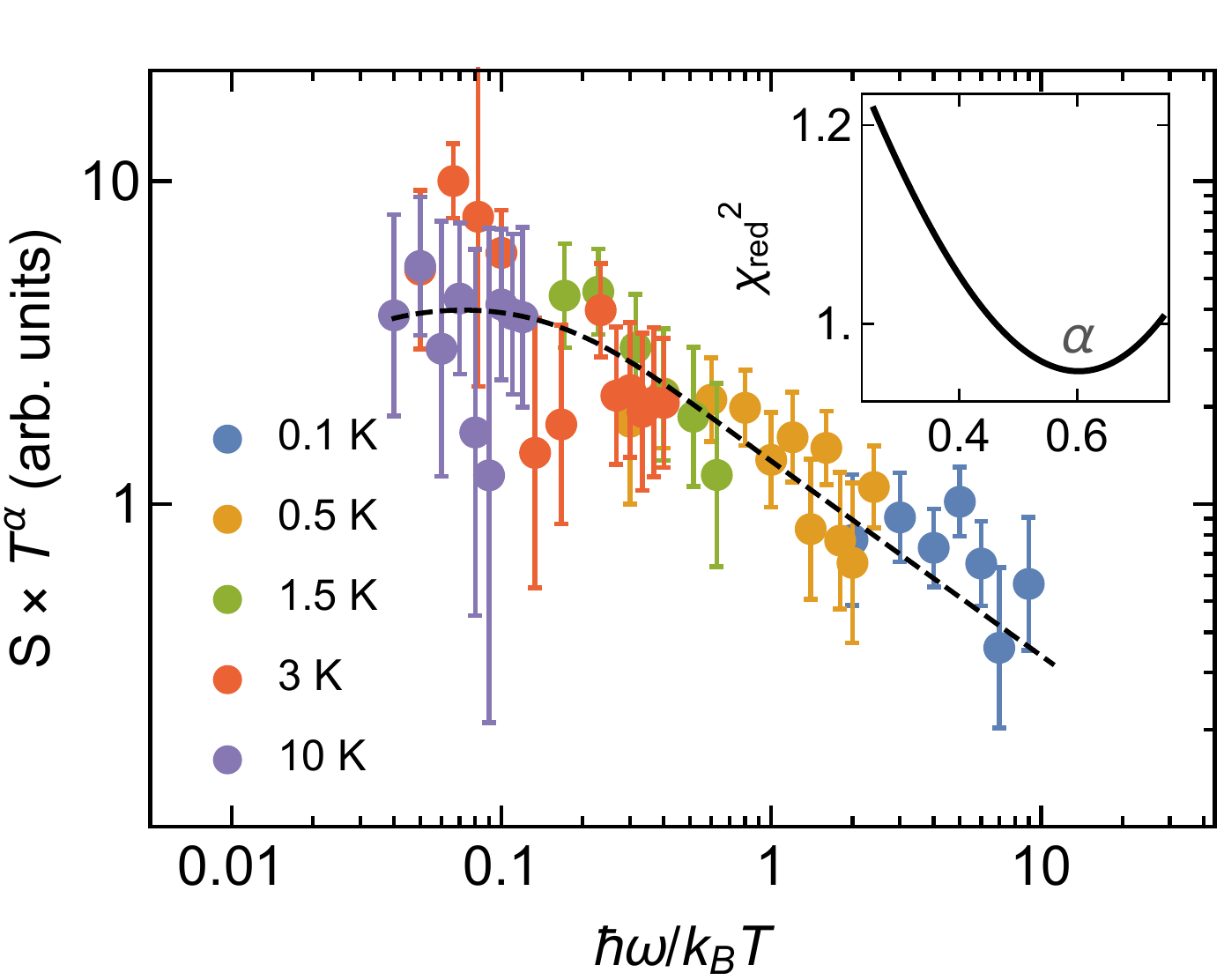}
\vspace{0cm}

\includegraphics[height=.3\textwidth]{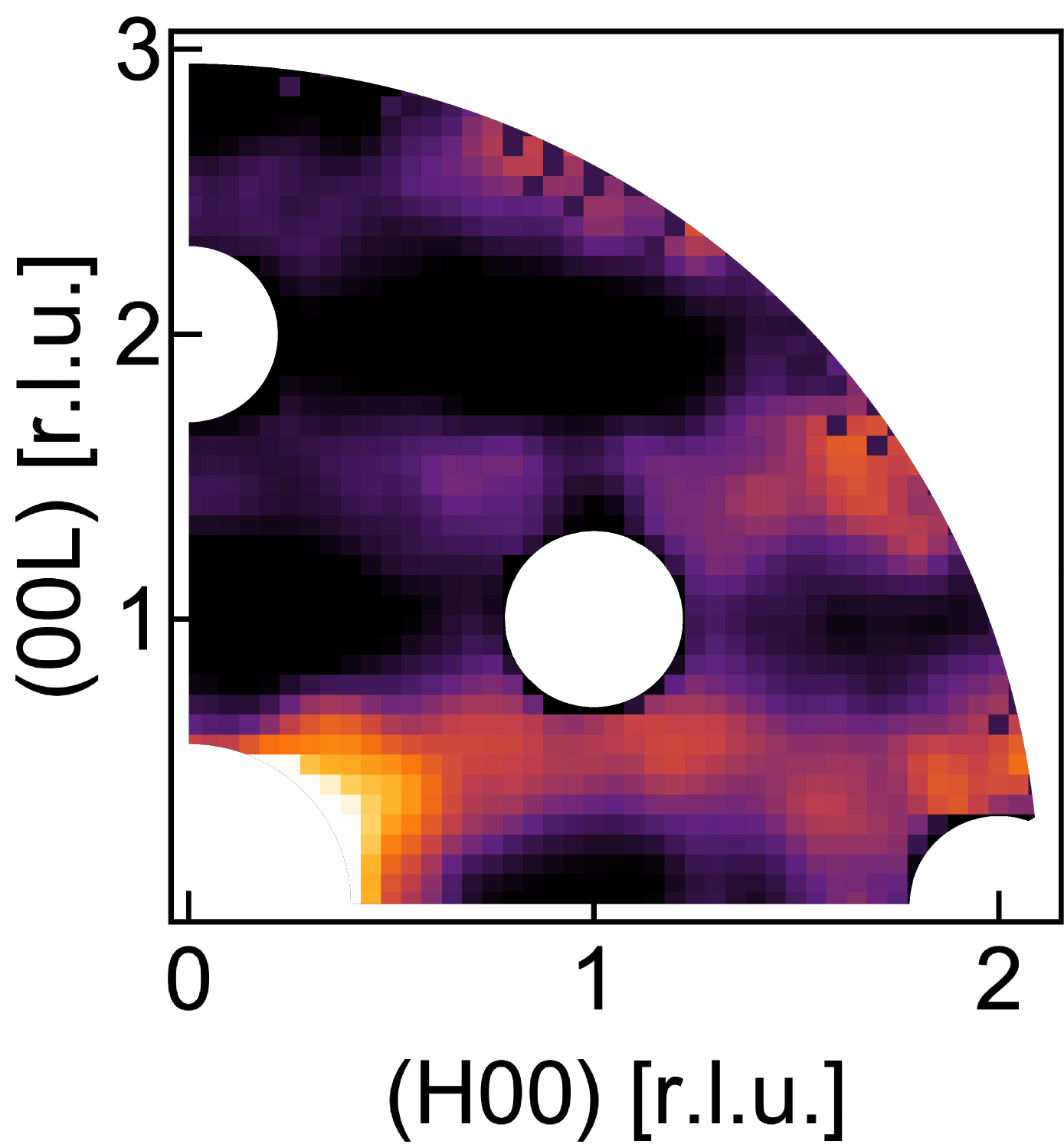}\hspace{0.3cm}\includegraphics[height=.3\textwidth]{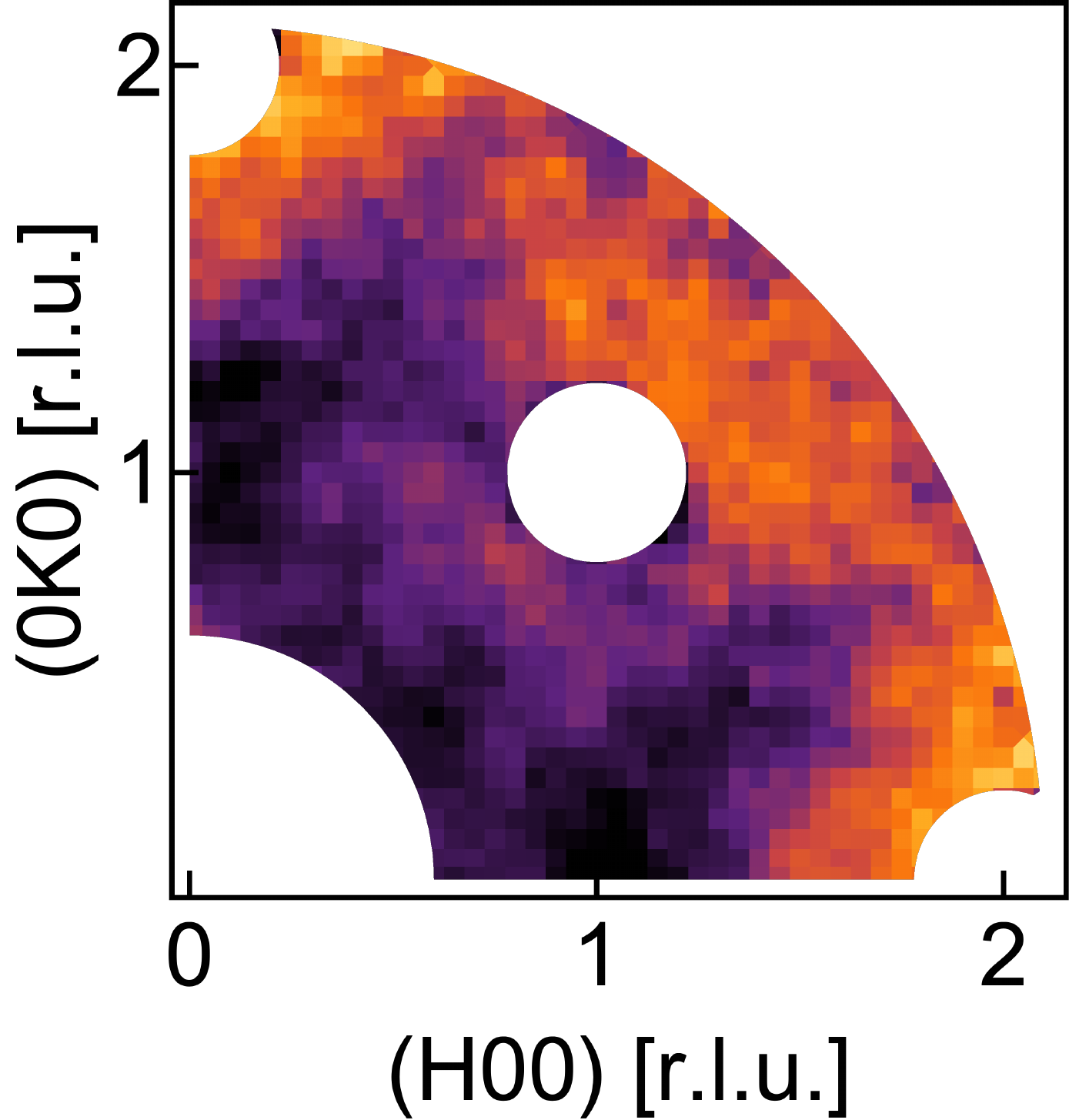}\hspace{0.3cm}\includegraphics[height=.35\textwidth]{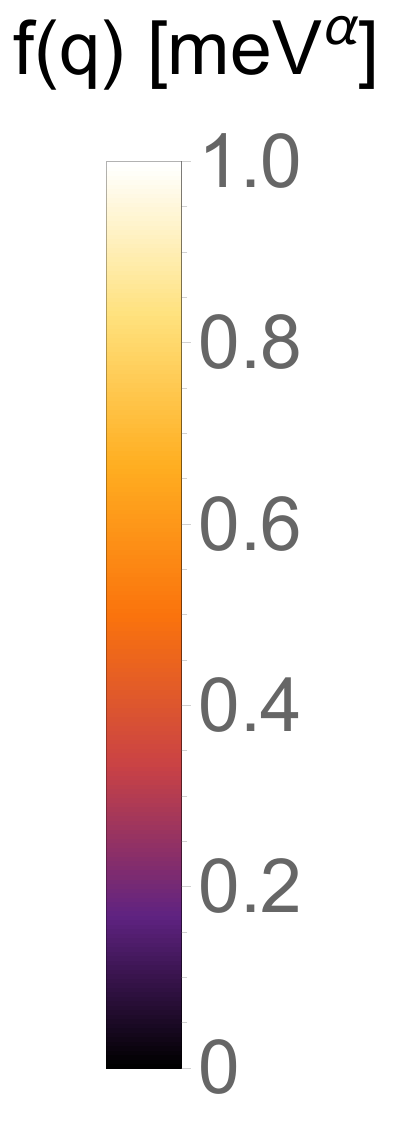}
\vspace{-15.2cm}

\hspace{-11.5cm}{\bf\large\textsf{a}}
\vspace{8.5cm}

\hspace{-6.5cm}{\bf\large\textsf{b}}\hspace{4.7cm}{\bf\large\textsf{c}}
\vspace{5.5cm}

\caption{\label{fig4}
{\bf Dynamical scaling of CeRu$_4$Sn$_6$.} {\bf (a)} Scaled neutron scattering
intensity, ${\cal S}\cdot T^{\alpha}$, vs scaled energy transfer,
$\hbar\omega/k_{\rm B}T$. The inset shows the scaled $\chi_{\rm red}^2$, which
is minimal for $\alpha = 0.6$. {\bf (b,c)} $f({\bf q})$ (see
Eqn.\,\ref{eq:chi-q-omega_FitFq}) in the $({\rm H0L})$ and $({\rm HK0})$ plane,
respectively, obtained through a global fit of Eqn.\,\ref{eq:chi-q-omega_FitFq}
to the $q$-$E$ dependent data in Fig.\,\ref{fig3} with a single consistent value
of $\alpha$. The white areas correspond to blinded out data near the nuclear
Bragg peaks, which are situated at ${\rm H\!\!+\!\!K\!\!+\!\!L\!=\!even}$. Along
$(\frac{1}{2}\!\!+\!\!{\rm H}\, \frac{1}{2}\!\!-\!\!{\rm H}\, 0)$ in {\bf c} and
about $({\rm H} 0 1)$ in {\bf b}, $f({\bf q})$ is minimal, indicating maximally
antiferromagnetic interactions (see text).}
\end{center}
\end{figure}

First, however, we investigate whether quantum critical scaling, as evidenced by
the static uniform magnetic susceptibility $\chi = (\partial M/\partial
H)_{H\rightarrow 0} = \chi'({\bf q}=0,\omega=0,T)$ (Fig.\,\ref{fig1}d,e, insets), is
also found in the dynamical spin susceptibility $\chi({\bf q},\omega,T)$. This
would give rise to the form
\begin{equation}
\chi({\bf q},\omega,T) = \frac{c}{f({\bf q})+(-i\hbar\omega + aT)^{\alpha}}\;,
\label{eq:chi-q-omega_FitFq}
\end{equation}
where $f({\bf q})$ is an offset and $\alpha = 1$ is expected for a Lorentzian fluctuation spectrum as
prescribed by the Landau order parameter description \cite{Her76.1,Mil93.1}, and
a fractional exponent $\alpha < 1$ points to its failure. The dynamic structure
factor ${\cal S}({\bf q},\omega,T)$ measured by inelastic neutron scattering is related
to the imaginary part of $\chi({\bf q},\omega,T)$ which, near the ordering wave vector, should take the form
\begin{equation}
\chi''(\omega,T) = \frac{1}{T^{\alpha}}\cdot g(\hbar\omega/k_{\rm B}T)\;.
\end{equation}
We have determined ${\cal S}(\omega,T) \sim \chi''(\omega,T)$ by integrating the
neutron scattering intensity for momentum transfers ${\bf q}$ about the $(100)$
wavevector, and plot it for temperatures between 0.1\,K and 10\,K, and energy
transfers between 0.1\,meV and 1.3\,meV as ${\cal S}\cdot T^{\alpha}$ vs $\hbar\omega/k_{\rm
B}T$ in Fig.\,\ref{fig4}a. The best data collapse is found for $\alpha =
0.6 \pm 0.1$ (minimum in $\chi^2$, inset of Fig.\,\ref{fig4}a). This is consistent with the $\alpha$ values determined from the static uniform magnetic
susceptibility 
($\alpha = 0.50 \pm 0.01$ for $H\perp c$ and $\alpha = 0.78 \pm
0.03$ for $H\parallel c$), in particular in view of the fact that the neutron
scattering intensity is a directional average, taken around the $(100)$ wavevector, where spin fluctuations polarized along and perpendicular to $c$ contribute in equal measures.
Thus, also the inelastic neutron scattering data evidence quantum criticality
beyond the long-wavelength fluctuations of an order parameter
\cite{Her76.1,Mil93.1}, such as expected in the theory of Kondo destruction
quantum criticality \cite{Si01.1}.

Finally, we turn to the spatial profile of the critical fluctuations. To do so,
we have fitted the  the experimental scattering intensity, corrected for the Ce
form factor,  in terms of Eqn.\,\ref{eq:chi-q-omega_FitFq}. The resulting
$f({\bf q})$ in the $({\rm H}0{\rm L})$ and $({\rm HK}0)$ planes are shown in 
Fig.\,\ref{fig4}b and c, respectively. For the rather featureless
$(\frac{1}{2}\!\!+\!\!{\rm H}\, \frac{1}{2}\!\!-\!\!{\rm H}\, 0)$  ``ridge'',
the singular dynamical spin susceptibility discussed earlier  corresponds to a
vanishingly small $f({\bf q})$ (see Fig.\,\ref{fig4}c). Likewise, in the $({\rm
H}0{\rm L})$ plane, it is seen that $f({\bf q})$ is vanishingly small for ${\rm
H\!\!+\!\!0\!\!+\!\!L\!=\!odd}$, where the dynamical spin  susceptibility is
also peaked. Along directions that move away from these ridges, the decreasing
dynamical spin susceptibility corresponds to an increasing $f({\bf q})$.  We can
understand these features of $f({\bf q})$ in terms of the wavevector dependence
of the  Ruderman-Kittel-Kasuya-Yosida (RKKY) interaction
\begin{equation}
J({\bf q}) = J_{\rm K}^2 L({\bf q},\omega=0) = J_{\rm K}^2 \sum_{\bf k} \frac{f(\epsilon_{\bf k}) - f(\epsilon_{\bf k+q})}{
 \epsilon_{\bf k+q} - \epsilon_{\bf k} } \; .
\end{equation}
Here, $J_{\rm K}$ is the Kondo coupling strength and $L({\bf q},\omega)$ the
Lindhard function of the (uncorrelated) conduction electrons, which we have
determined through LDA-DFT calculations, with one $4f$ electron per Ce$^{3+}$
ion placed in the ionic core. This ``$f$-core" bandstructure, resembling that of
LaRu$_4$Sn$_6$, characterizes the electronic structure of the $spd$ conduction
electrons that mediate the RKKY interaction between the $4f$ moments. The
associated Fermi surface is shown in Fig.\,\ref{fig3}f, from which we can
identify nesting wavevectors that  correspond to the $(\frac{1}{2}\!\!+\!\!{\rm
H}\, \frac{1}{2}\!\!-\!\!{\rm H}\, 0)$ and ${\rm H\!\!+\!\!0\!\!+\!\!L\!=\!odd}$
ridges. These can be seen as the broad planes perpendicular to the ${\rm(00L})$
direction ($\Gamma$ to $M$) and between $X$ points extending along ${\rm(00L})$.
At these wavevectors, collectively denoted as ${\bf Q}$, the RKKY interaction is
maximally antiferromagnetic. The Weiss temperature scale
\begin{equation}
\Theta({\bf q}) = J({\bf q}) - J({\bf Q})
\end{equation}
provides an understanding of the offset $f({\bf q})$, when the latter is raised
to the power $1/\alpha$. In other words, the neutron scattering intensity is
maximal where the Weiss temperature  $\Theta({\bf q})$ has a minimum, implying
that  the RKKY interaction defines the ${\bf q}$ space structure. Accordingly,
the system may be near antiferromagnetic order mediated by the RKKY interaction,
with an ordering wavevector at ${\rm H\!\!+\!\!K\!\!+\!\!L\!=\!odd}$.

Beyond-Landau quantum criticality, as indicated by a fractional exponent $\alpha
< 1$, has been documented in a handful of other heavy fermion compounds
\cite{Sch00.1,Ish02.1,Mat11.1,Bau08.2,Bau01.2}. Except for the case of the heavy
fermion metal CeCu$_{5.9}$Au$_{0.1}$, a determination of $\alpha$ from inelastic
neutron scattering---as provided here---has, however, remained elusive. Our
discovery of beyond-Landau quantum criticality, in both the inelastic neutron
response and the static uniform magnetic susceptibility, in a genuinely quantum
critical semimetal implicates the generality of the phenomenon. Since this
constitutes the first observation of this phenomenon in a semimetal, it will be
important to explore whether beyond-Landau quantum criticality is inherent to
systems with reduced charge carrier concentration. 

In heavy fermion systems, quantum criticality is generally observed at the $T=0$
collapse of (antiferromagnetically) ordered phases. CeRu$_4$Sn$_6$, however, is
quantum critical without tuning pressure, stoichiometry, or applied fields to a
phase boundary. A natural question then arises: Is an antiferromagnetic phase
nearby, as suggested by the minima of the Weiss temperature near possible
antiferromagnetic ordering wavevectors, and can it be reached by tuning, for
instance with pressure? And even more excitingly, do the quantum critical
fluctuations give rise to new emergent phases, maybe unconventional
superconductivity in analogy to $\beta$-YbAlB$_4$ (Ref.\citenum{Nak08.1}) or, in
view of all the necessary conditions being fulfilled, even a Weyl-Kondo
semimetal phase akin to that recently discovered in Ce$_3$Bi$_4$Pd$_3$ (refs 
\citenum{Dzs17.1,Lai18.1,Dzs18.1x})? This will require further experiments at
lower temperatures and as function of tuning parameters, which we hope this work
will stimulate.

\clearpage

\newpage


\clearpage

\newpage


\clearpage

\newpage


\clearpage

\newpage


\clearpage
\newpage


\noindent{\bf Methods}

\noindent For our study, single crystals of CeRu$_4$Sn$_6$ were grown from
self-flux, using the floating zone melting technique with optical heating as
reported previously \cite{Pas10.1}. The magnetic properties between 2\,K and
300\,K were measured in a SQUID magnetometer from Cryogenic Ltd., the data
between 0.3\,K and 2\,K employed a $^3$He insert. High-temperature magnetization
measurements up to 1000\,K were performed with a vibrating sample magnetometer
in a PPMS from Quantum Design. The $^3$He PPMS option was used for specific heat
measurements in the temperature range from 0.3\,K to 20\,K. 

Neutron scattering was performed with the MACS spectrometer at the NIST NCNR.
For the momentum-space mapping, we co-aligned 3\,g of single crystals and used 
incident energies from 3.7\,meV to 5\,meV and a fixed final energy of 3.7\,meV,
allowing for the use of Be and BeO filters before and after the sample,
respectively, and providing an effective resolution better than 0.05\,\AA$^{-1}$
and 0.15\,meV. For the $\omega/T$ scaling measurements, 2\,g of single crystals
were accommodated within the restrictions of a dilution refrigerator. We
utilized a fixed final energy of 2.5\,meV and incident energies up to 3.7\,meV.
A magnetic field of 10\,mT was applied along $(0 {\rm K} 0)$ to avoid
superconductivity in the aluminum sample mount and maintain good thermal
contact.\\

\noindent{\bf Acknowledgements}

\noindent The work at IQM was supported as part of the Institute for Quantum
Matter, an Energy Frontier Research Center funded by the U.S. Department of
Energy, Office of Science, Basic Energy Sciences under Award No. DE-SC0019331.
CB was supported by the Gordon and Betty Moore Foundation through GBMF-4532.
S.P. and A.S. acknowledge financial support from the European Community (H2020
Project No. 824109) and from the Austrian Science Fund (FWF projects P29296-N27,
29279-N27, and W1243).  Q.S. was supported by the NSF (Grant No. DMR-1920740) 
and the Robert A. Welch Foundation (Grant No. C-1411), and acknowledges the
hospitality of the Aspen Center for Physics, which is  supported by the NSF
(Grant No. PHY-1607611). WTF is grateful to the ARCS foundation, Lockheed Martin
and KPMG for the partial support of this work and to Predrag Nikolic for
fruitful discussions. 

\newpage


\begin{thebibliography}{10}
\expandafter\ifx\csname url\endcsname\relax
  \def\url#1{\texttt{#1}}\fi
\expandafter\ifx\csname urlprefix\endcsname\relax\def\urlprefix{URL }\fi
\providecommand{\bibinfo}[2]{#2}
\providecommand{\eprint}[2][]{\url{#2}}

\bibitem{Col10.1}
\bibinfo{author}{Coldea, R.}, \bibinfo{author}{Tennant, D.~A.},
  \bibinfo{author}{Wheeler, E.~M.}, \bibinfo{author}{Wawrzynska, E.},
  \bibinfo{author}{Prabhakaran, D.}, \bibinfo{author}{Telling, M.},
  \bibinfo{author}{Habicht, K.}, \bibinfo{author}{Smeibidl, P.} \&
  \bibinfo{author}{Kiefer, K.}
\newblock \bibinfo{title}{{Quantum criticality in an Ising chain: Experimental
  evidence for emergent E$_8$ symmetry}}.
\newblock \emph{\bibinfo{journal}{{Science}}} \textbf{\bibinfo{volume}{327}},
  \bibinfo{pages}{177} (\bibinfo{year}{2010}).

\bibitem{Sac10.1}
\bibinfo{author}{Sachdev, S.}
\newblock \bibinfo{title}{{Where is the quantum critical point in the cuprate
  superconductors?}}
\newblock \emph{\bibinfo{journal}{{Phys.\ Status Solidi B}}}
  \textbf{\bibinfo{volume}{247}}, \bibinfo{pages}{537} (\bibinfo{year}{2010}).

\bibitem{Dai09.1}
\bibinfo{author}{Dai, J.}, \bibinfo{author}{Si, Q.}, \bibinfo{author}{Zhu,
  J.-X.} \& \bibinfo{author}{Abrahams, E.}
\newblock \bibinfo{title}{{Iron pnictides as a new setting for quantum
  criticality}}.
\newblock \emph{\bibinfo{journal}{{Proc.\ Natl.\ Acad.\ Sci.\ U.S.A.}}}
  \textbf{\bibinfo{volume}{106}}, \bibinfo{pages}{4118} (\bibinfo{year}{2009}).

\bibitem{Loe07.1}
\bibinfo{author}{{v. L\"ohneysen}, H.}, \bibinfo{author}{Rosch, A.},
  \bibinfo{author}{Vojta, M.} \& \bibinfo{author}{W\"olfle, P.}
\newblock \bibinfo{title}{{Fermi-liquid instabilities at magnetic quantum
  critical points}}.
\newblock \emph{\bibinfo{journal}{{Rev.\ Mod.\ Phys.}}}
  \textbf{\bibinfo{volume}{79}}, \bibinfo{pages}{1015} (\bibinfo{year}{2007}).

\bibitem{Kir20.1}
\bibinfo{author}{Kirchner, S.}, \bibinfo{author}{Paschen, S.},
  \bibinfo{author}{Chen, Q.}, \bibinfo{author}{Wirth, S.},
  \bibinfo{author}{Feng, D.}, \bibinfo{author}{Thompson, J.~D.} \&
  \bibinfo{author}{Si, Q.}
\newblock \bibinfo{title}{Colloquium: Heavy-electron quantum criticality and
  single-particle spectroscopy}.
\newblock \emph{\bibinfo{journal}{Rev. Mod. Phys.}}
  \textbf{\bibinfo{volume}{92}}, \bibinfo{pages}{011002}
  (\bibinfo{year}{2020}).

\bibitem{Bit96.1}
\bibinfo{author}{Bitko, D.}, \bibinfo{author}{Rosenbaum, T.~F.} \&
  \bibinfo{author}{Aeppli, G.}
\newblock \bibinfo{title}{{Quantum critical behavior for a model magnet}}.
\newblock \emph{\bibinfo{journal}{{Phys.\ Rev.\ Lett.}}}
  \textbf{\bibinfo{volume}{77}}, \bibinfo{pages}{940} (\bibinfo{year}{1996}).

\bibitem{Her76.1}
\bibinfo{author}{Hertz, J.~A.}
\newblock \bibinfo{title}{{Quantum critical phenomena}}.
\newblock \emph{\bibinfo{journal}{{Phys.\ Rev.\ B}}}
  \textbf{\bibinfo{volume}{14}}, \bibinfo{pages}{1165} (\bibinfo{year}{1976}).

\bibitem{Mil93.1}
\bibinfo{author}{Millis, A.~J.}
\newblock \bibinfo{title}{{Effect of a nonzero temperature on quantum critical
  points in itinerant fermion systems}}.
\newblock \emph{\bibinfo{journal}{{Phys.\ Rev.\ B}}}
  \textbf{\bibinfo{volume}{48}}, \bibinfo{pages}{7183} (\bibinfo{year}{1993}).

\bibitem{Kue03.1}
\bibinfo{author}{K\"uchler, R.}, \bibinfo{author}{Oeschler, N.},
  \bibinfo{author}{Gegenwart, P.}, \bibinfo{author}{Cichorek, T.},
  \bibinfo{author}{Neumaier, K.}, \bibinfo{author}{Tegus, O.},
  \bibinfo{author}{Geibel, C.}, \bibinfo{author}{Mydosh, J.~A.},
  \bibinfo{author}{Steglich, F.}, \bibinfo{author}{Zhu, L.} \&
  \bibinfo{author}{Si, Q.}
\newblock \bibinfo{title}{{Divergence of the Gr\"uneisen ratio at quantum
  critical points in heavy fermion metals}}.
\newblock \emph{\bibinfo{journal}{{Phys.\ Rev.\ Lett.}}}
  \textbf{\bibinfo{volume}{91}}, \bibinfo{pages}{066405}
  (\bibinfo{year}{2003}).

\bibitem{Jar10.1}
\bibinfo{author}{Jaramillo, R.}, \bibinfo{author}{Feng, Y.},
  \bibinfo{author}{Wang, J.} \& \bibinfo{author}{Rosenbaum, T.~F.}
\newblock \bibinfo{title}{{Signatures of quantum criticality in pure Cr at high
  pressure}}.
\newblock \emph{\bibinfo{journal}{{Proc.\ Natl.\ Acad.\ Sci.\ U.S.A.}}}
  \textbf{\bibinfo{volume}{{107}}}, \bibinfo{pages}{{13631}}
  (\bibinfo{year}{{2010}}).

\bibitem{Sch00.1}
\bibinfo{author}{Schr\"{o}der, A.}, \bibinfo{author}{Aeppli, G.},
  \bibinfo{author}{Coldea, R.}, \bibinfo{author}{Adams, M.},
  \bibinfo{author}{Stockert, O.}, \bibinfo{author}{{v.\ L\"{o}hneysen}, H.},
  \bibinfo{author}{Bucher, E.}, \bibinfo{author}{Ramazashvili, R.} \&
  \bibinfo{author}{Coleman, P.}
\newblock \bibinfo{title}{{Onset of antiferromagnetism in heavy-fermion
  metals}}.
\newblock \emph{\bibinfo{journal}{{Nature}}} \textbf{\bibinfo{volume}{407}},
  \bibinfo{pages}{351} (\bibinfo{year}{2000}).

\bibitem{Si01.1}
\bibinfo{author}{Si, Q.}, \bibinfo{author}{Rabello, S.},
  \bibinfo{author}{Ingersent, K.} \& \bibinfo{author}{Smith, J.}
\newblock \bibinfo{title}{{Locally critical quantum phase transitions in
  strongly correlated metals}}.
\newblock \emph{\bibinfo{journal}{Nature}} \textbf{\bibinfo{volume}{413}},
  \bibinfo{pages}{804} (\bibinfo{year}{2001}).

\bibitem{Lee09.2}
\bibinfo{author}{Lee, S.-S.}
\newblock \bibinfo{title}{{Low-energy effective theory of Fermi surface coupled
  with U(1) gauge field in $2+1$ dimensions}}.
\newblock \emph{\bibinfo{journal}{Phys. Rev. B}} \textbf{\bibinfo{volume}{80}},
  \bibinfo{pages}{165102} (\bibinfo{year}{2009}).

\bibitem{Mro10.1}
\bibinfo{author}{Mross, D.~F.}, \bibinfo{author}{McGreevy, J.},
  \bibinfo{author}{Liu, H.} \& \bibinfo{author}{Senthil, T.}
\newblock \bibinfo{title}{{Controlled expansion for certain non-Fermi-liquid
  metals}}.
\newblock \emph{\bibinfo{journal}{Phys. Rev. B}} \textbf{\bibinfo{volume}{82}},
  \bibinfo{pages}{045121} (\bibinfo{year}{2010}).

\bibitem{Gur13.1}
\bibinfo{author}{Guritanu, V.}, \bibinfo{author}{Wissgott, P.},
  \bibinfo{author}{Weig, T.}, \bibinfo{author}{Winkler, H.},
  \bibinfo{author}{Sichelschmidt, J.}, \bibinfo{author}{Scheffler, M.},
  \bibinfo{author}{Prokofiev, A.}, \bibinfo{author}{Kimura, S.},
  \bibinfo{author}{Iizuka, T.}, \bibinfo{author}{Strydom, A.~M.},
  \bibinfo{author}{Dressel, M.}, \bibinfo{author}{Steglich, F.},
  \bibinfo{author}{Held, K.} \& \bibinfo{author}{Paschen, S.}
\newblock \bibinfo{title}{{Anisotropic optical conductivity of the putative
  Kondo insulator CeRu$_4$Sn$_6$}}.
\newblock \emph{\bibinfo{journal}{{Phys.\ Rev.\ B}}}
  \textbf{\bibinfo{volume}{87}}, \bibinfo{pages}{115129}
  (\bibinfo{year}{2013}).

\bibitem{Sun15.2}
\bibinfo{author}{Sundermann, M.}, \bibinfo{author}{Strigari, F.},
  \bibinfo{author}{Willers, T.}, \bibinfo{author}{Winkler, H.},
  \bibinfo{author}{Prokofiev, A.}, \bibinfo{author}{Ablett, J.~M.},
  \bibinfo{author}{Rueff, J.~P.}, \bibinfo{author}{Schmitz, D.},
  \bibinfo{author}{Weschke, E.}, \bibinfo{author}{{Moretti Sala}, M.},
  \bibinfo{author}{Al-Zein, A.}, \bibinfo{author}{Tanaka, A.},
  \bibinfo{author}{Haverkort, M.~W.}, \bibinfo{author}{Kasinathan, D.},
  \bibinfo{author}{Tjeng, L.~H.}, \bibinfo{author}{Paschen, S.} \&
  \bibinfo{author}{Severing, A.}
\newblock \bibinfo{title}{{CeRu$_4$Sn$_6$: a strongly correlated material with
  nontrivial topology}}.
\newblock \emph{\bibinfo{journal}{{Sci.\ Rep.}}} \textbf{\bibinfo{volume}{5}},
  \bibinfo{pages}{17937} (\bibinfo{year}{2015}).

\bibitem{Xu17.1}
\bibinfo{author}{Xu, Y.}, \bibinfo{author}{Yue, C.}, \bibinfo{author}{Weng, H.}
  \& \bibinfo{author}{Dai, X.}
\newblock \bibinfo{title}{{Heavy Weyl fermion state in CeRu$_4$Sn$_6$}}.
\newblock \emph{\bibinfo{journal}{Phys.\ Rev.\ X}}
  \textbf{\bibinfo{volume}{7}}, \bibinfo{pages}{011027} (\bibinfo{year}{2017}).

\bibitem{Dzs17.1}
\bibinfo{author}{Dzsaber, S.}, \bibinfo{author}{Prochaska, L.},
  \bibinfo{author}{Sidorenko, A.}, \bibinfo{author}{Eguchi, G.},
  \bibinfo{author}{Svagera, R.}, \bibinfo{author}{Waas, M.},
  \bibinfo{author}{Prokofiev, A.}, \bibinfo{author}{Si, Q.} \&
  \bibinfo{author}{Paschen, S.}
\newblock \bibinfo{title}{{Kondo insulator to semimetal transformation tuned by
  spin-orbit coupling}}.
\newblock \emph{\bibinfo{journal}{{Phys.\ Rev.\ Lett.}}}
  \textbf{\bibinfo{volume}{118}}, \bibinfo{pages}{246601}
  (\bibinfo{year}{2017}).

\bibitem{Lai18.1}
\bibinfo{author}{Lai, H.-H.}, \bibinfo{author}{Grefe, S.~E.},
  \bibinfo{author}{Paschen, S.} \& \bibinfo{author}{Si, Q.}
\newblock \bibinfo{title}{{Weyl-Kondo semimetal in heavy-fermion systems}}.
\newblock \emph{\bibinfo{journal}{{Proc.\ Natl.\ Acad.\ Sci.\ U.S.A.}}}
  \textbf{\bibinfo{volume}{115}}, \bibinfo{pages}{93} (\bibinfo{year}{2018}).

\bibitem{Dzs18.1x}
\bibinfo{author}{Dzsaber, S.}, \bibinfo{author}{Yan, X.},
  \bibinfo{author}{Eguchi, G.}, \bibinfo{author}{Prokofiev, A.},
  \bibinfo{author}{Shiroka, T.}, \bibinfo{author}{Blaha, P.},
  \bibinfo{author}{Rubel, O.}, \bibinfo{author}{Grefe, S.~E.},
  \bibinfo{author}{Lai, H.-H.}, \bibinfo{author}{Si, Q.} \&
  \bibinfo{author}{Paschen, S.}
\newblock \bibinfo{title}{{Giant spontaneous Hall effect in a nonmagnetic
  Weyl-Kondo semimetal. {\em arXiv:1811.02819} (2018)}}.

\bibitem{Mat11.1}
\bibinfo{author}{Matsumoto, Y.}, \bibinfo{author}{Nakatsuji, S.},
  \bibinfo{author}{Kuga, K.}, \bibinfo{author}{Karaki, Y.},
  \bibinfo{author}{Horie, N.}, \bibinfo{author}{Shimura, Y.},
  \bibinfo{author}{Sakakibara, T.}, \bibinfo{author}{Nevidomskyy, A.~H.} \&
  \bibinfo{author}{Coleman, P.}
\newblock \bibinfo{title}{{Quantum criticality without tuning in the mixed
  valence compound $\beta$-YbAlB$_4$}}.
\newblock \emph{\bibinfo{journal}{Science}} \textbf{\bibinfo{volume}{331}},
  \bibinfo{pages}{316} (\bibinfo{year}{2011}).

\bibitem{Amo18.1}
\bibinfo{author}{Amorese, A.}, \bibinfo{author}{Kummer, K.},
  \bibinfo{author}{Brookes, N.~B.}, \bibinfo{author}{Stockert, O.},
  \bibinfo{author}{Adroja, D.~T.}, \bibinfo{author}{Strydom, A.~M.},
  \bibinfo{author}{Sidorenko, A.}, \bibinfo{author}{Winkler, H.},
  \bibinfo{author}{Zocco, D.~A.}, \bibinfo{author}{Prokofiev, A.},
  \bibinfo{author}{Paschen, S.}, \bibinfo{author}{Haverkort, M.~W.},
  \bibinfo{author}{Tjeng, L.~H.} \& \bibinfo{author}{Severing, A.}
\newblock \bibinfo{title}{{Determining the local low-energy excitations in the
  Kondo semimetal CeRu$_4$Sn$_6$ using resonant inelastic x-ray scattering}}.
\newblock \emph{\bibinfo{journal}{{Phys.\ Rev.\ B}}}
  \textbf{\bibinfo{volume}{98}}, \bibinfo{pages}{081116}
  (\bibinfo{year}{2018}).

\bibitem{Zhu03.1}
\bibinfo{author}{Zhu, L.}, \bibinfo{author}{Garst, M.}, \bibinfo{author}{Rosch,
  A.} \& \bibinfo{author}{Si, Q.}
\newblock \bibinfo{title}{{Universally diverging Gr\"uneisen parameter and the
  magnetocaloric effect close to quantum critical points}}.
\newblock \emph{\bibinfo{journal}{{Phys.\ Rev.\ Lett.}}}
  \textbf{\bibinfo{volume}{91}}, \bibinfo{pages}{066404}
  (\bibinfo{year}{2003}).

\bibitem{Tok09.2}
\bibinfo{author}{Tokiwa, Y.}, \bibinfo{author}{Radu, T.},
  \bibinfo{author}{Geibel, C.}, \bibinfo{author}{Steglich, F.} \&
  \bibinfo{author}{Gegenwart, P.}
\newblock \bibinfo{title}{{Divergence of the magnetic Gr{\"u}neisen ratio at
  the field-induced quantum critical point in YbRh$_2$Si$_2$}}.
\newblock \emph{\bibinfo{journal}{{Phys.\ Rev.\ Lett.}}}
  \textbf{\bibinfo{volume}{102}}, \bibinfo{pages}{066401}
  (\bibinfo{year}{2009}).

\bibitem{Geg16.1}
\bibinfo{author}{Gegenwart, P.}
\newblock \bibinfo{title}{{Gr\"uneisen parameter studies on heavy fermion
  quantum criticality}}.
\newblock \emph{\bibinfo{journal}{{Rep.\ Prog.\ Phys.}}}
  \textbf{\bibinfo{volume}{79}}, \bibinfo{pages}{114502}
  (\bibinfo{year}{2016}).

\bibitem{Ish02.1}
\bibinfo{author}{Ishida, K.}, \bibinfo{author}{Okamoto, K.},
  \bibinfo{author}{Kawasaki, Y.}, \bibinfo{author}{Kitaoka, Y.},
  \bibinfo{author}{Trovarelli, O.}, \bibinfo{author}{Geibel, C.} \&
  \bibinfo{author}{Steglich, F.}
\newblock \bibinfo{title}{{YbRh$_2$Si$_2$: Spin fluctuations in the vicinity of
  a quantum critical point at low magnetic fields}}.
\newblock \emph{\bibinfo{journal}{{Phys.\ Rev.\ Lett.}}}
  \textbf{\bibinfo{volume}{89}}, \bibinfo{pages}{107202}
  (\bibinfo{year}{2002}).

\bibitem{Bau08.2}
\bibinfo{author}{Baumbach, R.}, \bibinfo{author}{Ho, P.},
  \bibinfo{author}{Sayles, T.}, \bibinfo{author}{Maple, M.},
  \bibinfo{author}{Wawryk, R.}, \bibinfo{author}{Cichorek, T.},
  \bibinfo{author}{Pietraszko, A.} \& \bibinfo{author}{Henkie, Z.}
\newblock \bibinfo{title}{{{Non-Fermi-liquid behavior in the filled
  skutterudite compound CeRu$_4$As$_{12}$}}}.
\newblock \emph{\bibinfo{journal}{{J.\ Phys.: Condens.\ Matter}}}
  \textbf{\bibinfo{volume}{20}}, \bibinfo{pages}{075110}
  (\bibinfo{year}{2008}).

\bibitem{Bau01.2}
\bibinfo{author}{Bauer, E.}, \bibinfo{author}{Slebarski, A.},
  \bibinfo{author}{Dickey, R.}, \bibinfo{author}{Freeman, E.},
  \bibinfo{author}{Sirvent, C.}, \bibinfo{author}{Zapf, V.},
  \bibinfo{author}{Dilley, N.} \& \bibinfo{author}{Maple, M.}
\newblock \bibinfo{title}{{Electronic and magnetic investigation of the filled
  skutterudite compound CeRu$_4$Sb$_{12}$}}.
\newblock \emph{\bibinfo{journal}{{J.\ Phys.: Condens.\ Matter}}}
  \textbf{\bibinfo{volume}{13}}, \bibinfo{pages}{5183} (\bibinfo{year}{2001}).

\bibitem{Nak08.1}
\bibinfo{author}{Nakatsuji, S.}, \bibinfo{author}{Kuga, K.},
  \bibinfo{author}{Machida, Y.}, \bibinfo{author}{Tayama, T.},
  \bibinfo{author}{Sakakibara, T.}, \bibinfo{author}{Karaki, Y.},
  \bibinfo{author}{Ishimoto, H.}, \bibinfo{author}{Yonezawa, S.},
  \bibinfo{author}{Maeno, Y.}, \bibinfo{author}{Pearson, E.},
  \bibinfo{author}{Lonzarich, G.}, \bibinfo{author}{Balicas, L.},
  \bibinfo{author}{Lee, H.} \& \bibinfo{author}{Fisk, Z.}
\newblock \bibinfo{title}{{Superconductivity and quantum criticality in the
  heavy-fermion system $\beta$-YbAlB$_4$}}.
\newblock \emph{\bibinfo{journal}{{Nature Phys.}}}
  \textbf{\bibinfo{volume}{4}}, \bibinfo{pages}{603} (\bibinfo{year}{2008}).

\bibitem{Ven90.1}
\bibinfo{author}{Venturini, G.}, \bibinfo{author}{{Chafik El Idrissi}, B.},
  \bibinfo{author}{Mar\^ech\'e, J.~F.} \& \bibinfo{author}{Malaman, B.}
\newblock \bibinfo{title}{{}}.
\newblock \emph{\bibinfo{journal}{{Mater.\ Res.\ Bull.}}}
  \textbf{\bibinfo{volume}{25}}, \bibinfo{pages}{1541} (\bibinfo{year}{1990}).

\bibitem{Das92.1}
\bibinfo{author}{Das, I.} \& \bibinfo{author}{Sampathkumaran, E.~V.}
\newblock \bibinfo{title}{{Electrical-resistance anomalies in the Ce-Ru-Sn
  phase}}.
\newblock \emph{\bibinfo{journal}{{Phys.\ Rev.\ B}}}
  \textbf{\bibinfo{volume}{46}}, \bibinfo{pages}{{4250}}
  (\bibinfo{year}{1992}).

\bibitem{Pas10.1}
\bibinfo{author}{Paschen, S.}, \bibinfo{author}{Winkler, H.},
  \bibinfo{author}{Nezu, T.}, \bibinfo{author}{Kriegisch, M.},
  \bibinfo{author}{Hilscher, G.}, \bibinfo{author}{Custers, J.},
  \bibinfo{author}{Prokofiev, A.} \& \bibinfo{author}{Strydom, A.}
\newblock \bibinfo{title}{{Anisotropy of the Kondo insulator CeRu$_4$Sn$_6$}}.
\newblock \emph{\bibinfo{journal}{{J.\ Phys.\ Conf.\ Series}}}
  \textbf{\bibinfo{volume}{200}}, \bibinfo{pages}{012156}
  (\bibinfo{year}{2010}).

\bibitem{Shi02.1}
\bibinfo{author}{Shirane, G.}, \bibinfo{author}{Shapiro, S.~M.} \&
  \bibinfo{author}{Tranquada, J.~M.}
\newblock \emph{\bibinfo{title}{{Neutron Scattering with a Triple-Axis
  Spectrometer}}} (\bibinfo{publisher}{Cambridge Univ.\ Press, Cambridge},
  \bibinfo{year}{2002}).

\bibitem{Xu96.1}
\bibinfo{author}{Xu, G.}, \bibinfo{author}{DiTusa, J.~F.},
  \bibinfo{author}{Ito, T.}, \bibinfo{author}{Oka, K.},
  \bibinfo{author}{Takagi, H.}, \bibinfo{author}{Broholm, C.} \&
  \bibinfo{author}{Aeppli, G.}
\newblock \bibinfo{title}{{Y$_2$BaNiO$_5$: A nearly ideal realization of the
  $S=1$ Heisenberg chain with antiferromagnetic interactions}}.
\newblock \emph{\bibinfo{journal}{{Phys.\ Rev.\ B}}}
  \textbf{\bibinfo{volume}{54}}, \bibinfo{pages}{R6827} (\bibinfo{year}{1996}).

\end{thebibliography}

\newpage

\noindent{\large\bf Supplementary Information}

\noindent{\bf Scaling and exponent relationship}

\noindent Here we derive a relationship between the scaling exponents, which we first summarize.
The exponent $\alpha$ characterizes the temperature dependence of the magnetization, and is defined as the exponent in
the power-law temperature dependence of the linear-response magnetic susceptibility
\begin{equation}
\chi(T) \sim \frac{1}{T^{\alpha}} \: .
\label{eq:chi}
\end{equation}
We also introduce $\beta$ and $\gamma$, which describe the magnetic-field dependence of the nonlinear magnetization. 
The exponent $\gamma$ is
defined by generalizing Eqn.\,\ref{eq:chi} to the case when the magnetic field is not vanishingly small, namely by
\begin{equation}
\frac{M}{H} = \frac{1}{T^{\alpha}}
 f(T/H^{\gamma})
\label{eq:MoverH}
\end{equation}
which, when taking the limit $H \rightarrow 0$, recovers Eqn.\,\ref{eq:chi}.
Additionally, the exponent $\beta$ is defined through the scaling of the
temperature derivative of the magnetization as
\begin{equation}
\frac{\partial (M/H)}{\partial T}  \times H^{\beta} = g(T/H^{\gamma})
\; .
\label{eq:gamma}
\end{equation}

Based on scaling (dimensional analysis), Eqn.\,\ref{eq:MoverH} implies that
\begin{equation}
\frac{\partial (M/H)}{\partial T} = \frac{1}{T^{1+\alpha} }
 f_1(T/H^{\gamma}) \; ,
\label{eq:d_MoverH_dT_T}
\end{equation}
which is equivalent to
\begin{equation}
\frac{\partial (M/H)}{\partial T} = \frac{1}{H^{\gamma(1+\alpha)} }
 g(T/H^{\gamma}) \; .
\label{eq:d_MoverH_dT_H}
\end{equation}

Comparing Eqs.\,\ref{eq:gamma} and \ref{eq:d_MoverH_dT_H} gives rise to  the
relationship
\begin{equation}
\beta = \gamma(1+\alpha)
\label{eq:exponent-relationship}
\end{equation}
among the three exponents.
 
We now apply this relationship to the exponents we have extracted for
CeRu$_4$Sn$_6$. Consider first $H \perp c$. The fitted exponents are
$\alpha=0.5$, $\gamma=0.35$, and $\beta=0.5$. The product $\gamma (1+\alpha) =
0.525$ is  compatible with $\beta$ within about $5$\% (thus within the experimental error bars, see main text). For $H \parallel c$, the fitted exponents are  $\alpha=0.78$, $\gamma=0.43$, and $\beta=0.78$. The product $\gamma (1+\alpha) = 0.765$
is consistent with $\beta$ within about $2$\%, again within experimental accuracy.

We note that we have assumed hyperscaling in the derivation of the scaling
relationship \cite{Zhu03.1}. The hyperscaling is expected to be satisfied at
interacting critical points. By contrast,  Gaussian critical points will contain
dangerously irrelevant variables that may invalidate the hyperscaling for  some
observables. The fact that the fitted exponents obey the scaling relationship is
consistent with the interacting  nature of the quantum critical point that
underlies the observations reported in the present work for CeRu$_4$Sn$_6$.\\

\noindent{\bf Neutron Scattering}

\noindent Background from the sample environment was accounted for by subtraction of a linear combination of separate measurements of count rates for an empty sample can and a CYTOP-only standard (CYTOP is a fluorine-based epoxy that we used to secure the CeRu$_4$Sn$_6$ single crystals). 

For incident energies below 3.7 meV, neutrons are incapable of undergoing Bragg diffraction with Al in and around the beam path, dramatically reducing the background associated with Bragg scattering from the split-coil magnet system. 

Sample-out background scattering was subtracted to remove contamination at low
scattering angles. To account for incoherent elastic nuclear scattering from the
sample, a pseudovoigt function was fit to the high-Q incoherent elastic profile
and scaled to the elastic line intensity with same magnitude of $Q$ as where our
$E/T$ scaling fit was performed, at a point in the Brillouin zone where magnetic
intensity is at a minimum. The data for $E/T$ scaling was collected at $(100)$,
and integrated $\pm 0.2$\,r.l.u.\ along $({\rm H} 0 0)$ and $(0 0 {\rm L})$.

The modulation in intensity outside of the first zone by translation of a reciprocal lattice vector provides information on the microscopic origin of the scattering. 
The magnetic form factor, $F_{\rm mag}$, modulates scattering intensity as $I(\bm q+ \bm G) = I(\bm{q}) \cdot |F_{\rm mag}(\bm{q}+\bm{G})|^2$ (Ref.\citenum{Shi02.1}). 
The response of unpolarized neutrons is then described by a linear combination of the components of the dynamic spin correlation function \cite{Xu96.1} as 
\begin{equation}
I ({\bf Q}, \omega) = \sin^2{\theta}\cdot {\cal S}^{aa}({\bf Q}, \omega) + (1-\sin^2\theta \cos^2\phi){\cal S}^{bb}({\bf Q}, \omega) + (1-\sin^2\theta \sin^2\theta){\cal S}^{cc}({\bf Q}, \omega) \; .
\end{equation}
We find that ${\cal S}^{aa}={\cal S}^{bb} = 2{\cal S}^{cc}$ improves the correspondence with the observed intensity modulation. This modulation effectively accounts for a magnetic form factor which is the result of anisotropic hybridization and/or easy plane crystal field anisotropy. 
Previous field-dependent heat capacity measurements show a field-induced extra contribution below about 10\,K, that is much larger for $H\perp c$ than for $H \parallel c$ (Ref.\citenum{Pas10.1}), consistent with our observation of a form factor and polarization factor which implies quasielastic fluctuations of easy-plane Ce $4f$ electrons at low temperatures.


\end{document}